%% file: main.tex
\documentclass[opre,nonblindrev]{informs3_arxiv}
\usepackage{stackengine}
\definecolor{green}{rgb}{1,0.5,0}
\usepackage{fancyhdr}

\OneAndAHalfSpacedXII % Current default line spacing
\usepackage{natbib}
 \bibpunct[, ]{(}{)}{,}{a}{}{,}%
 \def\bibfont{\small}%
 \def\bibsep{\smallskipamount}%
\usepackage{booktabs,tablefootnote}
\usepackage{xcolor}
\usepackage{hyperref}
\usepackage{thmtools}
\usepackage[noabbrev, capitalise]{cleveref}
\usepackage{footmisc}
\usepackage{bm}

\usepackage{algorithm}%http://ctan.org/pkg/algorithms
\usepackage{algpseudocode}%

\theoremstyle{plain}
\newtheorem{theorem}{Theorem}[section]

\newtheorem{lemma}[theorem]{Lemma}
\newtheorem{proposition}[theorem]{Proposition}

\newtheorem{definition}[theorem]{Definition}

\newcommand{\R}[0]{{\mathbb{R} }}

\newcommand{\vx}[0]{{\mathbf{x}}}
\newcommand{\vf}[0]{{\mathbf{f} }}
\newcommand{\vv}[0]{{\mathbf{v} }}
\newcommand{\vu}[0]{{\mathbf{u} }}

\newcommand{\vy}[0]{{\mathbf{y} }}

\DeclareSymbolFont{symbolsC}{U}{txsyc}{m}{n}
\DeclareMathSymbol{\notniFromTxfonts}{\mathrel}{symbolsC}{61}

\newcommand{\opt}{\mbox{\sf OPT}}

\DeclareMathOperator*{\E}{\mathbb{E}}

\usepackage{color-edits}
\addauthor[Fang-Yi]{fang}{green}
\addauthor[Amin]{mar}{blue}
\addauthor[Yuxin]{lyx}{red}
\addauthor[Space Save]{ss}{yellow}

\usepackage{xcolor}

\newcommand{\defn}[1]{{\textbf{\textit{#1}}}}

\newcommand{\qedwhite}{\hfill \ensuremath{\Box}}

\newcommand{\BibTeX}{\rm B\kern-.05em{\sc i\kern-.025em b}\kern-.08em\TeX}

\begin{document}

\RUNAUTHOR{Liu, Rahimian, and Yu}

\RUNTITLE{Seeding with Differentially Private Network Information}

\TITLE{Seeding with Differentially Private Network Information}

% \ARTICLEAUTHORS{%
% \AUTHOR{}
% \AFF{ \EMAIL{}}
% % \AUTHOR{}
% % \AFF{ \EMAIL{}}
% % \AUTHOR{}
% % \AFF{ \EMAIL{}}
% % \AUTHOR{}
% % \AFF{ \EMAIL{}}
% % Enter all authors
% \vspace{-30pt}
% } % end of the block

\ARTICLEAUTHORS{%
\AUTHOR{Yuxin Liu$^{1}$, M. Amin  Rahimian$^{1}$, Fang-Yi Yu$^{2}$}
\AFF{$^{1}$Industrial Engineering, University of Pittsburgh \quad $^{2}$Computer Science, George Mason University\\ \EMAIL{yul435@pitt.edu, rahimian@pitt.edu, fangyiyu@gmu.edu}}
}

\ABSTRACT{\input{sections/abstract}
}%Erd\H{o}s-R\'enyi network


\maketitle

%%% The following commands remove the headers in your paper. For final 
%%% papers, these will be inserted during the pagination process.

\pagestyle{fancy}
\fancyhead{}

%%% The next command prints the information defined in the preamble.

\maketitle 

%%%%%%%%%%%%%%%%%%%%%%%%%%%%%%%%%%%%%%%%%%%%%%%%%%%%%%%%%%%%%%%%%%%%%%%%

\input{sections/body}

% \textcolor{blue}{extensions to individual level, multiple agents and group privacy}

% \textcolor{blue}{Finally, although $L_2$ regularization has been proved empirically to be a simple, but powerful framework, to optimize the response of the Randomize response algorithm. We expect to expand the analysis to other families of regularizations and other network parameter settings both theoretically and empirically.}
%\clearpage
 
\bibliographystyle{informs2014} 
\bibliography{reference}


\begin{APPENDICES}
\renewcommand\thefigure{\thesection.\arabic{figure}} 
\setcounter{figure}{0}    
\phantomsection
\input{sections/appendix}

\end{APPENDICES}

% \newpage
\ACKNOWLEDGMENT{Liu, Rahimian, and Yu contributed equally and are listed alphabetically. This research was supported in part by NSF grant DMS-2424684 and by the University of Pittsburgh Center for Research Computing and Data, RRID:SCR\_022735, through the resources provided. Specifically, this work used the H2P cluster, which is supported by the NSF award number OAC-2117681. Data and code for reported simulations can be accessed from \url{https://github.com/aminrahimian/dp-inf-max}. A preliminary version of this work was presented at the 22nd International Conference on Autonomous Agents and Multiagent Systems (AAMAS 2023). The AAMAS proceedings includes a two-page extended abstract under the title, ``Differentially private network data collection for influence maximization'' \citep{rahimian2023differentially}. The present paper substantially extends that proceedings version by developing its informal results into a full theoretical and algorithmic treatment, including formal theorem statements and proofs, central and local privacy-preserving seeding algorithms with sample-complexity guarantees, and an empirical evaluation on MSM sexual-contact networks motivated by HIV prevention.}
%  The authors thank Juba Ziani for valuable discussions and feedback.}

\end{document}

%% file: sections/abstract.tex
In public health interventions such as distributing preexposure prophylaxis (PrEP) for HIV prevention, decision makers often use seeding algorithms to identify key individuals who can amplify intervention impact. However, building a complete sexual activity network is typically infeasible due to privacy concerns. Instead, contact tracing can provide \emph{influence samples}—observed sequences of sexual contacts—without full network reconstruction. This raises two challenges: protecting individual privacy in these samples and adapting seeding algorithms to incomplete data. We study differential privacy guarantees for influence maximization when the input consists of randomly collected cascades. Building on recent advances in costly seeding, we propose privacy-preserving algorithms that introduce randomization in data or outputs and bound the privacy loss of each node. Theoretical analysis and simulations on synthetic and real-world sexual contact data show that performance degrades gracefully as privacy budgets tighten, with central privacy regimes achieving better trade-offs than local ones.

%% file: sections/body.tex
\section{Introduction}
When designing interventions in public health, development and education, decision makers rely on data from social networks to target a small number of people, leveraging peer effects and social contagion to deliver the greatest welfare benefits to the population. Many of these seeding methods are implemented without complete knowledge of social network structures---for example, by targeting nominated friends of random individuals \citep{kim2015social} or random pairs of friends \citep{alexander2022algorithms}---and enhance adoption of public health interventions by driving indirect treatment effects among people who were affected by spillovers from others' treatment (even though they were not part of the initial treatment group). However, even without full knowledge of the underlying network, individuals who participate in partial social network data collection—especially members of minority or stigmatized groups—still face privacy risks.  {For instance, in the context of HIV intervention, if contact tracing begins with individuals who have already tested positive, others may be reluctant to participate in the survey due to the perceived risk of being associated with high-risk contacts—such as being mistakenly identified as part of a transmission chain, facing social stigma, or having sensitive health information inferred from their network connections.} At the same time, social network information plays a critical role in designing effective HIV prevention strategies \citep{Wilder_Onasch-Vera_Diguiseppi_Petering_Hill_Yadav_Rice_Tambe_2021}. For example, the distribution of pre-exposure prophylaxis (PrEP) can benefit from seeding algorithms that identify high-risk individuals based on data from social networks \citep{ross2016implementation}. However, constructing a complete sexual activity network for this purpose is challenging due to the sensitivity of the sexual behavior data. Instead, contact tracing can provide influence samples, which capture sequences of sexual activity without requiring full network knowledge. This introduces two key challenges: protecting the privacy of individuals contributing to contract tracing data (i.e., influence samples) and adapting seeding algorithms to effectively leverage these partial data. Developing privacy-preserving methods for network data collection and targeted interventions is critical for responsible public health interventions in social networks, particularly in vulnerable communities such as homeless youth, who are estimated to have a six to twelve times higher risk of HIV than housed youth \citep{caccamo2017narrative}. Such methods ensure that privacy concerns do not undermine the trust and
community engagement necessary for successful public health initiatives, complementing
broader data-driven approaches to infectious-disease intervention under limited information,
such as screening under scarce testing resources \citep{el2022screening}
and spatiotemporal vaccine allocation under behavioral feedback dynamics
\citep{barth2024spatiotemporal}.

% When designing interventions in public health, development, and education, decision makers rely on data from social networks to target a small number of people, leveraging peer effects and social contagion to deliver the greatest welfare benefits to the population. Many of these seeding methods are implemented without complete knowledge of social network structures -- e.g., by targeting nominated friends of random individuals \citep{kim2015social} or random pairs of friends \citep{alexander2022algorithms} -- and they enhance adoption of public health interventions by driving indirect treatment effects among people who were affected by spillovers from others' treatment (even though they were not part of the initial treatment group). However, even without complete network information, network participants (and in particular minority members) who contribute to partial social network data collection endure privacy risks. This is especially important when vulnerable populations are involved in recording sensitive or stigmatizing behavior, for example, in the design of public health interventions among homeless youth for HIV or suicide prevention \citep{Wilder_Onasch-Vera_Diguiseppi_Petering_Hill_Yadav_Rice_Tambe_2021}. Developing new methods that preserve privacy for network data collection and targeted interventions is critical to responsible design of public health interventions on social networks and maintaining community engagements that support such interventions.

In this paper, we use differential privacy to provide participants with
plausible deniability about their presence in the collected influence samples.
The privacy parameter \(\epsilon\), often called the privacy budget, controls the
strength of this protection: smaller \(\epsilon\) gives stronger privacy, but
requires more randomization and may reduce targeting efficacy. We study this
privacy--utility trade-off in two regimes. In the local regime, sample entries
are randomized before submission. In the central regime, a trusted curator
observes the samples and randomizes the final seed-selection output.

%In \Cref{sec:contributions}, we elaborate on shortcomings of mainstream approaches to DP analysis of graphs for seeding and use that to contextualize our main contributions in proposing new DP definitions with privacy, utility, and sample complexity guarantees for influence maximization.  

% In \Cref{sec:contributions}, we elaborate on the mainstream approaches to social network data privacy and explain their shortcomings for seeding interventions to contextualize our main contributions in proposing new privacy definitions with utility and sample complexity guarantees for seeding social network interventions. In \Cref{sec:related-work}, we provide a more in-depth overview of literature on both seeding and differential privacy in the network diffusions context. Next we provide preliminaries in  \Cref{sec:pre}, including hardness results for seeding with node- and edge-DP guarantees, as well as our proposed notion of influence sample differential privacy. In \Cref{sec:mechanisms}  we address both of the preceding privacy regimes, central DP in \Cref{sec:exp} followed by local DP in \Cref{sec:rr}. In \Cref{sec:sim}, we test the performance of our algorithms (local and central DP) at different values of privacy budget ($\epsilon$) and sample size ($m$) on empirical social network datasets. We end with a discussion of limitations, future directions and concluding remarks in \Cref{sec:conc}. Complete proofs, additional simulations and supplementary results are provided in several appendix sections.

After summarizing our main contributions in \cref{sec:contributions} and providing detailed comparisons with related works in \cref{sec:related-work}, in \cref{sec:pre} we provide preliminaries including hardness results that demonstrate the inadequacy of mainstream approaches to the privacy of social network data for seeding interventions and motivate our proposed notion of ``influence sample differential privacy''. In \Cref{sec:mechanisms}, we provide privacy mechanisms for seeding in central (\Cref{sec:exp}) and local (\Cref{sec:rr}) regimes. In \Cref{sec:sim}, we test the performance of our privacy mechanisms on empirically-grounded networks of men who have sex with men (MSM). We end with a discussion of limitations, future directions, and the concluding remarks in \Cref{sec:conc}. Complete proofs and supplementary results are provided in several appendices at the end. % additional simulations 
\subsection{Summary of the main results} \label{sec:contributions}

Differential privacy (DP) offers an information-theoretic foundation for privacy by focusing on protecting input data against revelations of algorithmic outputs \citep{dwork2014algorithmic}. In mathematical formulation, a randomized mechanism $\mathcal{M}$ is $\epsilon$-DP if for any two adjacent input datasets $D$ and $D'$ in the input domain and any subset $A$ in the output range we have:
    \begin{align}
        \frac {\Pr[\mathcal{M}(D) \in A]} {\Pr [\mathcal{M}(D') \in A]} \le e^\epsilon.
        \label{eq:DP-def}
    \end{align}    
The $\epsilon$-DP criterion in \cref{eq:DP-def} limits the inferences of any observer/adversary about the input data based on the algorithm output. The notion of adjacency between datasets is typically defined so that neighboring datasets differ in contributions of an individual entity, and thus $\epsilon$-DP protects individuals' contributions from ramification of an adversarial inference. Hence, different notions of adjacency imply different privacy protections. 
In our setting, the input is not the full network but a collection of partial
diffusion records, which we call \emph{influence samples}. We therefore define
adjacency at the level of these samples: two inputs are adjacent if they differ
in one individual's presence in one sample. This adjacency relation determines
the privacy protection studied in the rest of the paper.

\paragraph{Seeding with partial and private network information.} We study a classical, NP-hard network optimization problem, known as ``influence maximization'', which asks to select $k$ nodes on a graph of size $n$ such that the expected spread size from seeding those $k$ nodes is maximum in a randomized network diffusion model, whereby each edge passes on the influence with an independent cascade probability $p$. Given the social network graph $G$ and an integer $k\ge 1$, we denote the expected spread size from infecting the nodes in $S$ by $I_G(S)$ and write $\opt = \max_{S:|S| \le k} I_G(S)$ for the optimum solution to the \emph{$k$-influence maximization problem}. It is an example of a cardinality-constrained combinatorial optimization on a graph with a monotone, submodular objective that admits a tight $(1-1/e)$ approximation guarantee using the greedy node selection algorithm. This problem has occupied a central place in the graph and social networks algorithms literature since its introduction by \citet{kempe2003maximizing}. There is a wealth of literature (spanning almost two decades) on efficient approximations of this problem \citep{kempe2003maximizing,Kempe2015-pi,chen2009efficient,borgs2014maximizing, du2024adaptive}. However, the graph structure of the input data has stunned the design of DP influence maximization algorithms. Common operationalizations of DP for graph algorithms, such as node or edge DP, define adjacent inputs in \Cref{eq:DP-def} based on neighboring graphs differing in a node or an edge. However, this adjacency criterion is often too stringent for network-based problems whose outputs are highly sensitive to such structural changes. Even the removal or addition of one node or edge can substantially alter the entire contagion process, leading to drastic changes in the resulting diffusion outcomes (see \cref{sec:related-work} and \Cref{sec:app:ISDP_Ext} for detailed citations and comparisons). Our main hardness results indicate that it is impossible to obtain useful performance guarantees for influence maximization subject to node- or edge-DP constraints (see \Cref{sec:ISDP}, \Cref{prop:node:dp:neg,prop:edge:dp:neg}).

\paragraph{A new privacy notion for social network data collection and intervention design.}
Our main contribution is to formulate a new differential privacy notion—
\emph{influence-sample differential privacy (ISDP)}—for influence maximization
when the full social network is unknown and the algorithm operates on partial
diffusion records rather than the entire network graph. We call these records
\emph{influence samples}. Informally, one influence sample corresponds to one
realization of a diffusion process and records which nodes could reach a sampled
target node through the realized diffusion paths. Thus, an influence sample can
be represented as a binary vector over the node set: an entry is one if the
corresponding node can reach the sampled target in that realization, and zero
otherwise. A collection of such samples provides partial information about
diffusion in the underlying network without requiring the planner to observe or
reconstruct the full graph.

This abstraction is useful in settings such as contact tracing or behavioral
surveys, where the data collector may observe partial records of who is connected
to whom through a particular activity, event, or exposure pathway, but cannot
observe the entire social network. Building on prior work on seeding with costly
network information \citep{costlyseeding} and the sample complexity of influence
maximization \citep{sadeh2020sample,borgs2014maximizing}, our framework protects
the privacy of individuals whose behavioral data are used for intervention
design.

\paragraph{Influence Sample Differential Privacy (ISDP).}
We define \emph{\(\epsilon\)-influence-sample differential privacy}
(\(\epsilon\)-ISDP) by controlling how much the distribution of the seeding
algorithm's output can change when one entry of the influence-sample data is
modified. Equivalently, two influence-sample datasets are adjacent if they differ
only in whether a single node appears in a single influence sample.

Influence samples operate at the \emph{activity level}: they record participation
or reachability in particular diffusion-related events, and may therefore reveal
whether a person had direct or indirect contact with high-risk individuals. This
activity-level notion of privacy can be understood as a form of
\emph{record-level differential privacy}, in which each
participation record is treated as an independent privacy unit
\citep{liu2024cross,messing2state}. Under \(\epsilon\)-ISDP, the privacy
parameter \(\epsilon\) bounds how much the released seed set can depend on any
one individual's presence in any one influence sample. Smaller values of
\(\epsilon\) correspond to stronger privacy guarantees.

To illustrate the type of information that ISDP protects, consider a contact-tracing scenario during a disease outbreak such as COVID-19. Investigators may identify several social gatherings or public events—such as parties, concerts, or community meetings—as potential starting points of infection. Each tracing record then corresponds to one such event and indicates who attended it, forming a partial view of how the infection might have spread through direct or indirect contacts. While these data are essential for designing effective interventions, they are also highly sensitive: participation in a single event record could reveal a person’s association with potentially infected individuals or their other political, personal or professional orientations that are associated with the event. An adversary observing the final intervention targets could thus infer whether someone appeared in these event records. ISDP mitigates this risk by bounding the influence of any single event record on the seeding outcome, ensuring that adding or removing one person’s participation in an event does not substantially alter the recommended targets. In doing so, ISDP gives individuals the ability to \emph{plausibly deny} their attendance at high-risk events, while still enabling accurate population-level intervention design.

\paragraph{Sample complexity with differential privacy and approximation guarantees.} 
We analyze ISDP under two standard privacy regimes—\emph{central} and \emph{local} differential privacy—which differ in the level of trust placed in the curator. Under \emph{central DP}, a trusted curator accesses the raw cascade data and enforces privacy by perturbing the algorithm’s output. Under \emph{local DP}, individuals randomize their own influence samples before submission, thereby protecting privacy at the data-collection stage. 

To achieve these guarantees, we design two efficient $\epsilon$-ISDP seeding algorithms based on the \emph{exponential mechanism}~\citep{mcsherry2007mechanism} and the \emph{randomized response mechanism}~\citep{warner1965randomized}, corresponding to the central and local privacy regimes, respectively. We derive performance bounds in terms of sample size $m$, network size $n$, seed set size $k$, and privacy budget $\epsilon$. Our results show that the central $\epsilon$-ISDP algorithm achieves near-optimal influence, infecting at least $(1 - 1/e)\opt - \alpha n$ nodes with high probability, and requires about 
$m = \max\{\tfrac{12}{\alpha\epsilon}, \tfrac{9}{\alpha^2}\}k \ln n$ 
influence samples. In contrast, the local $\epsilon$-ISDP algorithm needs substantially more samples due to the additional noise introduced by local randomization.(see details in \cref{sec:mechanisms}, \Cref{thm:exp:acc,thm:rr:acc}).

\subsection{Comparisons with related work} \label{sec:related-work}

\paragraph{Seeding network diffusions.} 
A long line of literature, going back to the seminal works of \citet{kempe2003maximizing,Kempe2015-pi}, proposes efficient approximation algorithms for influence maximization based on the submodularity of the influence function \citep{chen2009efficient,wang2012scalable}.This problem has been widely applied in marketing and public health to promote the adoption of new products and interventions through social networks \citep{han2023cost, kahr2024impact}. For example, \citet{banerjee2013diffusion} demonstrate that network-based seeding substantially improves participation in social programs, while \citet{kim2015social} show that targeting key individuals can increase community adoption of water purification methods in field experiments. In the context of public health, social network interventions have been used to increase awareness of HIV prevention among homeless youth \citep{yadav2018bridging} and to reduce bullying among adolescents \citep{paluck2016changing}. In all such settings, collecting complete social network data is difficult, costly, or infeasible, and poses critical privacy risks to individuals. In this paper, we focus on providing differential privacy guarantees for influence maximization when the social network is unknown and influential nodes must be inferred from limited and potentially sensitive diffusion samples.

\paragraph{Social network data privacy.} Much of the previous work on social network data privacy focuses on the identifiability of nodes from released graph data, with or without additional information on node attributes and their neighborhoods \citep{romanini2021privacy}. Classical anonymization techniques mask the attributes of the nodes and perturb, modify, randomize, or aggregate the structure of the graph to prevent reidentification of the nodes within a confidence level, typically achieving $k$ anonymity \citep{wu2010survey,zheleva2011privacy}. %\citep{wu2010survey,zheleva2011privacy,abawajy2016privacy}. 
However, statistical disclosure control of collected data requires safeguarding data donor's privacy against adversarial attacks where common anonymization techniques are shown to be vulnerable to various kinds of identification \citep{barbaro2006face}, linkage and cross-referencing \citep{narayanan2008robust,sweeney2015only}, statistical difference and reidentification attacks \citep{kumar2007anonymizing}. However, while edge sampling at data collection can somewhat mitigate the risk of reidentification of nodes \citep{romanini2021privacy,jiang2021applications}, seeded nodes can reveal additional side information about who has contributed to data collection. For example, once a community leader is chosen, his or her close contacts may suffer privacy leaks even though no social network data is published. Our privacy guarantees for influence maximization follow the DP formalism to provide network participants and data donors with plausible deniability with regard to their contribution to the collected data or their role in shaping the algorithm output (the seeded nodes). DP is a gold standard for privacy-preserving data analysis and avoids the above drawbacks by abstracting away from the attack models and controlling the privacy leak at an information-theoretic level according to \cref{eq:DP-def}.

\paragraph{DP analysis of graphs.} 
Differential privacy (DP) provides a general framework for designing privacy-preserving algorithms and has been adopted in large-scale applications such as the US Census \citep{garfinkel2022differential} and major technology platforms \citep{erlingsson2014learning,appleDP}. 
It has been successfully applied to a wide range of optimization and learning problems, including combinatorial \citep{gupta2010differentially}, submodular \citep{pmlr-v70-mitrovic17a}, online submodular optimization \citep{salazar2021differentially}, and personalized statistical learning under output perturbation \citep{acharya2026personalized}.
Existing work on differential privacy for graphs—such as edge DP \citep{nissim2007smooth,eden2023triangle} and node DP \citep{kasiviswanathan2013analyzing,raskhodnikova2016lipschitz,kalemaj2023node}—has primarily focused on protecting global graph statistics or subgraph counts \citep{blocki2022privately,hay2009accurate}. 
However, these approaches are not directly applicable to influence maximization, and our hardness results show that node or edge DP is often too stringent to yield meaningful performance guarantees. 
Similar limitations have been observed for other graph algorithms that are highly sensitive to the addition or removal of individual nodes or edges \citep{gupta2010differentially,epasto2022differentially}. To address these limitations, we introduce \emph{influence sample differential privacy} (ISDP), which shifts the input from an explicit social network to a collection of randomly sampled diffusion cascades. 
ISDP bounds information leakage about each node’s participation in these cascades—either at the data level (local ISDP) or through the algorithm’s output (central ISDP). While local ISDP can be achieved by randomizing individual participation reports, coordinating such randomization in a decentralized and context-aware manner remains challenging \citep{imola2021locally,dhulipala2022differential}.

\section{Problem Setup and Preliminary Results}\label{sec:pre}
\subsection{Influence sampling and maximization} \label{sec:subsec:IMP}

In the independent cascade (IC) model, diffusion takes place on an
edge-weighted undirected graph \(G=(V,p)\) with \(n\) nodes. In our motivating
public-health setting, nodes represent individuals in the target population,
and an edge represents a potential contact pathway between two individuals.
The weight \(p_e=p(v,u)=p(u,v)\in[0,1]\) on edge \(e=(v,u)\) represents the
probability that this pathway successfully transmits the intervention effect
in one diffusion instance.

Given a seed set \(S\subseteq V\), the diffusion process starts from the nodes in
\(S\). These are the individuals initially selected for an intervention, such as
receiving PrEP outreach, prevention information, or another targeted treatment.
The process then unfolds in discrete rounds. Each newly reached node has one
opportunity to activate each of its neighbors, and edge \(e\) succeeds with
probability \(p_e\). Thus, the model captures the idea that an intervention can
generate spillovers beyond the initially selected individuals through social or
contact pathways.

Following \citet{kempe2003maximizing}, this stochastic diffusion process can be
viewed equivalently as sampling a random realization of successful diffusion
links. Specifically, we draw an unweighted graph \(g\) from \(G\) by including
each edge \(e\) independently with probability \(p_e\). The graph \(g\) is a realized version of the diffusion process, indicating
which potential contact pathways would succeed in that particular instance.
Conditional on \(g\), the spread from any seed set \(S\) is deterministic: the
eventual set of reached nodes is exactly the set of nodes reachable from \(S\)
in \(g\). This reachability-based formulation offers a convenient way to analyze
influence propagation and will be used throughout the paper.

\paragraph{Influence function.}
For a realized diffusion graph \(g\) and a seed set \(S\subseteq V\), let
\(C_g(S)\) denote the set of nodes that can be reached from \(S\) through paths
in \(g\). In the intervention setting, \(C_g(S)\) is the group of individuals
who would eventually be reached by the intervention in this particular diffusion
instance, starting from the initially selected individuals \(S\). We define the
realized influence of \(S\) as
\[
    I_g(S)=|C_g(S)|,
\]
the number of individuals reached in realization \(g\).

Since \(g\) is random, the realized influence \(I_g(S)\) may vary across
diffusion instances. The expected influence of \(S\) under the IC model is
\[
    I_G(S):=\mathbb{E}_{g\sim \mathcal{G}}[I_g(S)],
\]
where \(\mathcal{G}\) denotes the distribution over live-edge realizations
induced by \(G\). Thus, \(I_G(S)\) measures the average intervention reach of
seed set \(S\).

We also use marginal influence notation. For two seed sets \(S,T\subseteq V\),
define
\[
    C_g(S\mid T):=C_g(S)\setminus C_g(T).
\]
This is the set of nodes reached by \(S\) in realization \(g\) that would not
already be reached by \(T\). Its expected size is
\[
    I_G(S\mid T):=\mathbb{E}_{g\sim \mathcal{G}}\big[|C_g(S\mid T)|\big].
\]
This quantity captures the additional expected reach obtained by adding \(S\)
on top of the nodes already reached by \(T\).
 
\paragraph{The influence maximization problem.}
Given graph $G$ and integer $k\ge 1$, the \emph{$k$-influence maximization problem} is to find a set $S\subseteq V$ of at most $k$ nodes maximizing the expected number of reachable nodes, $I_G(S)$. 
We write $\opt = \max_{S:|S| \le k} I_G(S)$ for the maximum expected influence of any set of $k$ nodes.

\paragraph{Influence samples under limited information.}
While there exists extensive literature on developing fast algorithms for
maximizing influence when the contagion model \(G=(V,p)\) is known, in many
real-world settings the underlying network and edge probabilities are
unobservable. Learning \(G\) directly may be too costly or infeasible because of
privacy restrictions and incomplete contact data. We therefore consider a query
model in which the algorithm does not observe the full contact network. Instead,
it observes \emph{influence samples}, which are partial records of reachability
generated from the diffusion process.

In our motivating HIV-prevention setting, an influence sample can be interpreted
as a limited tracing record centered on one sampled individual. The record does
not reveal the full sexual-contact network. Rather, it records which individuals could be connected
to the sampled individual through the contact pathways. Multiple samples may come from repeated tracing
or survey records centered on different sampled individuals. The underlying population \(V\) and
diffusion model \(G\) are fixed, but the sampled individual and the realized
diffusion pathways vary across samples.

Formally, to generate one influence sample, we first sample a target node
\(u\in V\) uniformly at random and draw a graph realization \(g\sim \mathcal G\),
where \(\mathcal G\) is the distribution over realized diffusion graphs induced
by \(G\). The algorithm does not observe \(g\) itself. It observes only a binary
vector
\[
    x=(x_1,\ldots,x_n)^\top\in\{0,1\}^n,
    \qquad
    x_v=\mathbf 1\{u\in C_g(v)\}.
\]
Thus, \(x_v=1\) means that node \(v\) can reach the sampled target \(u\) in the
realization \(g\). In a contact-tracing interpretation, this entry records that
\(v\) is connected to the target individual \(u\) through the realized exposure
pathways captured by this sample. For example, \(v\) and \(u\) may have attended
the same gathering, visited the same location during a relevant time window, or
been connected through a short chain of contacts identified by tracing. Equivalently, we also view \(x\) as the subset \(x=\{v\in V: u\in C_g(v)\}\), 
namely the set of nodes that could reach the sampled target in this realization. 
When \(x_v=1\), we say that node \(v\) appears in the influence sample \(x\).

A collection of \(m\) independent influence samples is denoted by
\[
    \vx=(x^1,\ldots,x^m)\in\{0,1\}^{n\times m}.
\]
Each \(x^t\) is generated by independently sampling a target node \(u_t\) and a
graph realization \(g_t\sim\mathcal G\), for \(t=1,\ldots,m\). In the contact
tracing example, these samples can be viewed as multiple tracing records
centered on different sampled individuals and exposure instances. We use \([m]\) as
shorthand for the index set \(\{1,\ldots,m\}\).

For simplicity, we identify each influence sample $x$ with its corresponding set of active nodes and use standard set notation. For any subset $S\subseteq V$, $S\cap x$ denotes the set of nodes in $S$ that appear in the sample $x$. When referring to a single node $v\in V$, we write $S\cap v$ as shorthand for $S\cap\{v\}$. This convention will be used throughout the paper.

\paragraph{Influence sample greedy seeding algorithm.} How can we design a seeding algorithm $\mathcal{M}$ that takes $m$ influence samples $\vx$ and produces a set of $k$ nodes?  Here we present a seeding algorithm by \citet{costlyseeding}.  The core idea of \Cref{alg:original} is to estimate the influence function from the influence samples and to run the greedy algorithm. Specifically, given $m$ realized influence samples $\vx$ and a subset of nodes $S$ in $G$, we define the \emph{$S$-reachable influence samples} as: $C_\vx(S) := \{t\in[m]: \exists v\in S, x_v^t = 1\},$ which is the set of influence samples whose activated nodes are reachable from the set $S$, and $C_\vx(S|T) = C_\vx(S)\setminus C_\vx(T)$. Alternatively, if we treat $\vx$ as a collection of $m$ subsets $x^1, \dots, x^m$, $C_\vx(S) = \{t:S\cap x^t\neq \emptyset\}$.  Finally, we define $I_\vx(S) = \frac{n}{m}|C_\vx(S)|$ and $I_\vx(S|T) = \frac{n}{m}|C_\vx(S|T)|$.  The following proposition states that the expectation of $I_\vx$ is equal to $I_{G}$.  We will use $v$ instead of $\{v\}$ for singleton sets to simplify the notation, for example, writing $I_G(v)$ for $I_G(\{v\})$.

\begin{proposition}[Observation 3.2 of \citet{DBLP:journals/corr/abs-1212-0884} -- cf. \citet{borgs2014maximizing}]
\label{prop:inf_ch}
Given a positive integer \(m\), a graph \(G=(V,p)\), and sets of nodes
\(S,T\subseteq V\), let \(\vx=(x^1,\ldots,x^m)\in\{0,1\}^{n\times m}\) be
\(m\) independent influence samples generated as defined above. Then
\(\E_{\vx}[I_{\vx}(S)] = I_G(S)\) and
\(\E_{\vx}[I_{\vx}(S\mid T)] = I_G(S\mid T)\). Additionally, by Hoeffding's
inequality, for all \(\epsilon>0\),
\(\Pr[|I_{\vx}(S)-I_G(S)|\ge \epsilon n]\le 2\exp(-2\epsilon^2m)\), where the
probability is over the randomness in the influence-sampling procedure.
\end{proposition}

With the above observation, \citet{costlyseeding} design a seeding algorithm that only uses influence samples and show that if the number of influence samples is of the order $k^2(\log n/\alpha)/\alpha^3$, then \Cref{alg:original} is optimal up to the additive error $\alpha n$. Informally, \Cref{alg:original} is a greedy algorithm that sequentially queries \Cref{alg:original_single} for the seed that maximizes marginal improvement.  In this paper, we will follow a similar idea that combines a greedy algorithm and an estimation algorithm for marginal improvement.
 
{\small 
\begin{algorithm}
\caption{Influence sample greedy seeding algorithm}\label{alg:original}
  \begin{algorithmic}[1]
    \Require  $m$ influence samples $\vx$ and total number of seeds $k$.
    \State Set $S_0 = \emptyset$
    \For{$i$ from $1$ to $k$}
    \State $v_i = \mathcal{M}(\vx, S_{i-1})$ by \Cref{alg:original_single}\label{iter:greedy}
    \State $S_i = S_{i-1}\cup v_i$
    \EndFor
  \State Return $S_k$
  \end{algorithmic}
\end{algorithm}

\begin{algorithm}
\caption{Single seed selection, $\mathcal{M}(\vx, S)$}\label{alg:original_single}
  \begin{algorithmic}[1]
    \Require $m$ influence samples $\vx$ and a seed set $S$.
    % \State Discard the influence samples that intersect with already chosen seeds $\Lambda^*$
    \For{$v\in [n]\setminus S$} \Comment{Chooses the $|S|+1$-th seed.}
    \State Compute $I_\vx(v|S) = \frac{n}{m}|C_\vx(S\cup v)\setminus C_\vx(S)|$.
    \EndFor
    \State Return $\argmax_{v\in [n]\setminus S} I_\vx(v|S)$
  \end{algorithmic}
\end{algorithm}
}

In \Cref{alg:original}, we modify the baseline influence-sample-based greedy algorithm to sequentially call \Cref{alg:original_single} for selecting each seed. Instead of collecting fresh influence samples at every greedy step, \Cref{alg:original} reuses the existing influence samples, thereby achieving substantially better sample efficiency. We formally establish this improvement in \Cref{thm:exp:acc}.

\begin{theorem}[label = thm:original, name = Theorem 2 of \citet{costlyseeding}]
Let \(G=(V,p)\) be an IC instance with \(n\ge 2\) nodes, and let \(k\le n\).
For any \(0<\alpha\le 1\), \Cref{alg:original} uses at most
\(81k^2\log(6nk/\alpha)/\alpha^3\) influence samples and outputs a seed set
\(S\subseteq V\) with \(|S|\le k\) such that
\(I_G(S)\ge (1-1/e)\opt-\alpha n\).
\end{theorem}

The factor \((1-1/e)\) is the classical worst-case approximation guarantee of
the greedy algorithm for monotone submodular maximization under a cardinality
constraint. It provides the benchmark for the influence-sample greedy procedure
of \citet{costlyseeding}. Their result shows that, even without observing the
full graph \(G\), the greedy benchmark can be approached using sufficiently many
influence samples. In this paper, we ask whether comparable guarantees can be
obtained when privacy constraints are imposed on the influence-sample data.
\citet{costlyseeding} also show that one cannot hope to obtain a constant-factor
approximation guarantee using \(o(n)\) influence samples.

\begin{theorem}[Theorem 3 of \citet{costlyseeding}]\label{thm:sample_neg} 
Let $0< \alpha < 1$ be any constant.  For any algorithm with $o(n)$ samples, there exists a weighted graph $G$ so that the output of the algorithm can only infect fewer than $\alpha\opt$ nodes.
\end{theorem}

\subsection{Influence Sample Differential Privacy}\label{sec:ISDP}
Intuitively, a differentially private (DP) algorithm ensures that the algorithms output on two \emph{adjacent inputs} are close to each other, making them difficult to distinguish.  
% \begin{definition}[DP]
% Given a set $\mathcal{X}$ with an adjacency relationship $\sim$, a discrete output space $\mathcal{Y}$, and $\epsilon\ge 0$, a function $\mathcal{M}$ is $\epsilon$ differentially private (DP) if for all adjacent pair $x, x'\in \mathcal{X}$ with $x\sim x'$ and output $\vy\in \mathcal{Y}$,
% $$\Pr[\mathcal{M}(x) = \vy]\le e^\epsilon \Pr[\mathcal{M}(x') = \vy].$$
% \end{definition}
However, there are multiple alternative definitions of the adjacency relation in graph-structured input data.  One classical notion is \emph{node DP}, where two graphs are adjacent if they are different by the connections of one node.  Another is \emph{edge DP}, where two graphs are adjacent if they differ by one edge~\citep{hay2009accurate}.  {While these definitions are well established, they often require adding excessive noise to ensure privacy. Moreover, they are ill-suited for sensitive domains such as public health contact tracing, where the underlying network is unclear and constructing an explicit network may be infeasible. In practice, contact tracing relies on surveys from confirmed cases rather than on full network data, both for practical and privacy reasons. Given the contextual and dynamic nature of contact data, a fixed network representation is often poorly defined. We therefore abstract away from the network and work directly with \emph{influence samples}, which capture how contact patterns propagate over time without requiring complete network information. A more detailed discussion of the relationship between node- or edge-level DP and ISDP is provided in \Cref{sec:app:ISDP_Ext}.}

% {\color{red}While these definitions are well established, they often require adding excessive noise to ensure privacy. However, they are ill-suited for sensitive domains such as public health contact tracing, where the underlying network is unclear and constructing an explicit network may be infeasible.\fang{inaccurate.  Node DP does not require perturbation on the graph, e.g., our reduction in proposition B1.  The flow of ideas could be clearer.} In practice, contact tracing relies on surveys from confirmed cases rather than full network data, both for practicality and privacy reasons. Given the contextual and dynamic nature of contact data, a fixed network representation is poorly defined. We therefore abstract away from the network and work directly with \emph{influence samples}, which capture how contact patterns propagate over time without requiring full network information. A more detailed discussion is provided in \Cref{sec:app:ISDP_Ext}.}

We study algorithms that only use these influence samples, where two collections of $m$ influence samples, $\vx$ and $\vx'$ in $\{0,1\}^{n\times m}$, are \emph{``influence sample'' adjacent} if one entry can be modified to transform $\vx$ into $\vx'$. We denote this adjacency by $\vx \underset{is}{\sim} \vx'$. Given this adjacency relation on influence samples, we define a stochastic seeding algorithm $\mathcal{M}$ to be DP if the algorithm's output does not change significantly when applied to adjacent influence samples. {In the context of HIV prevention, the seed set output by the algorithm—corresponding to individuals recommended for pre-exposure prophylaxis (PrEP)—is typically observable in practice, because healthcare workers must implement the intervention for those selected individuals. \citet{doi:10.1073/pnas.1510612113} note that when performing social network search algorithms, it is important to distinguish between entities that require privacy protection and those that do not. The seed set, which represents the actionable output of the algorithm, falls into the latter category: it must be disclosed to enable implementation and thus cannot itself be protected under differential privacy. Instead, the privacy guarantee should apply to the process of identifying the seed set—specifically, to individuals’ participation in diffusion samples and their network relationships. Consequently, an adversary may still attempt to infer, from differences between adjacent influence samples, whether a specific individual is reachable from a high-risk participant, potentially revealing sensitive contact information and exposing them to social stigma or discrimination. %{\color{blue}solved}
}

%In this paper, we would like to focus on DP seeding algorithms that use influence samples but not raw graph data. Instead of a raw graph, we study algorithms that only use historical cascade data, that is, influence samples, where two collections of $m$ influence samples $\vx$ and $\vx'$ in $\{0,1\}^{n\times m}$ are \emph{``influence sample'' adjacent} if we can change one entry to convert $\mathbf{x}$ to $\mathbf{x}'$, and write $\vx\underset{is}{\sim} \vx'$ if they are adjacent. Given the above adjacency relation on influence samples, we can define if a stochastic seeding algorithm $\mathcal{M}$ is DP.

\vspace{-10pt}
\begin{definition}[ISDP]\label{def:ISDP}
Given $\epsilon\ge 0$, $n,m\ge 1$, and $k\in [n]$, a function $\mathcal{M}^k:\{0,1\}^{n\times m}\to  \mathcal{Y}$ is $\epsilon$-\defn{influence sample differentially private} ($\epsilon-$ISDP, in short) if for all outputs $\vy \in  \mathcal{Y}$ of $k$ seeds, and all pairs of adjacent datasets (collection of influence samples), $\mathbf{x} \underset{is}{\sim} \mathbf{x}'$, we have: $\Pr[\mathcal{M}^k(\mathbf{x}) = \vy]\le e^\epsilon \Pr[\mathcal{M}^k(\mathbf{x}') = \vy].$
\end{definition}

%\mar{should we say something about $(\epsilon,\delta)-$DP?}\fang{We can still have similar lower bounds that is added in the appendix.  We can allude to that but we may not need to introduce additional notions in the main body}

We believe that the study of influence sample DP is more natural and practical than conventional node and edge DP for seeding algorithms \citep{kasiviswanathan2013analyzing,raskhodnikova2016lipschitz}. First, the notion of node differential privacy (DP) allows an agent to misreport or withhold her social connections. However, in the context of network cascade data, an individual has limited ability to alter her network ties to affect the contagion dynamics; instead, she can decide whether to acknowledge or deny participation in a particular activity, such as denying that she visited a specific place in a contact-tracing survey. Therefore, we argue that defining adjacent datasets based on influence samples as input, rather than raw social network data, is more natural and realistic for contact tracing and public health applications.  Moreover, we show in \Cref{prop:node:dp:neg,prop:edge:dp:neg} that node DP and edge DP are too strict for influence maximization and hardly admit any utility guarantees.

Given two edge-weighted undirected graphs $G = (V, p)$ and $G' = (V', p')$, we say $G$ and $G'$ are \emph{node adjacent} if $V = V'$ and there exist $v\in V$, such that $p_e = p_e'$ for all $e \notniFromTxfonts v$.  Similarly, $G$ and $G'$ are \emph{edge adjacent} if $V = V'$ and there exists $u,v\in V$ such that $p_e = p_e'$ for all $e\neq (u,v)$.  For $k<n$, we say that a seeding algorithm $\mathcal{M}^k$ is $\epsilon$-node (or edge) DP if for any $k$, node (or edge) adjacent graphs $G, G'$, and output $\vy$, $\Pr[\mathcal{M}^k(G) = \vy]\le e^\epsilon \Pr[\mathcal{M}^k(G') = \vy]$.

The following proposition (proved in \Cref{proof:prop:node:dp:neg}) shows that seeding algorithms with node DP guarantees cannot produce high-quality seeds.

% \begin{proposition}[label = prop:node:dp:neg, name=Hardness of Influence Maximization with Node DP]
% Given $n\ge 1$ and $\epsilon>0$, for any $\epsilon$ node-DP seeding algorithm $\mathcal{M}^{k}(G)$, there exists $G$ of $n$ nodes so that for $k=1$ we have: $\textstyle \max_{S:|S| = 1}I_G(S)-\E_{\vy\sim \mathcal{M}^{k=1}(G)}[I_G(\vy)]\ge \frac{n}{4}-(e^\epsilon+1).$
% \end{proposition}
{%\color{red}
% \vspace{-10pt}
% \begin{proposition}[Hardness of Influence Maximization under Node DP]\label{prop:node:dp:neg}
% For any $\epsilon$-node DP seeding algorithm $\mathcal{M}^{k}(G)$ with $n>1$ and $\epsilon>0$, there exist graphs $G$ on $n$ nodes for which the following holds. When the maximum degree of $G$ is $n-1$, $\max_{S:|S|=1} I_G(S) - \E_{\vy\sim \mathcal{M}^{k=1}(G)}[I_G(\vy)] \ge \frac{n}{4} - (e^{\epsilon}+1)$. When the maximum degree satisfies $2\le d < n-1$,  $(1-1/e)\max_{S:|S|=1} I_G(S) - \E_{\vy\sim \mathcal{M}^{k=1}(G)}[I_G(\vy)] \ge\frac{1}{10e^{\epsilon}}
% \Bigl(1-\frac{d}{d-1}
% \Bigl(\frac{1}{e}+\frac{1}{10}\Bigr)\Bigr)n
% -\Bigl(1
% -\frac{d}{d-1} \Bigl(\frac{1}{e}+\frac{1}{10}\Bigr)\Bigr)\Bigl(1+\frac{1}{10e^{\epsilon}}\Bigr)$.
% \end{proposition}}
%\fang{

\begin{proposition}[Hardness of Influence Maximization under Node DP]\label{prop:node:dp:neg}
For any $\epsilon>0, n>1$, and $\epsilon$-node DP seeding algorithm $\mathcal{M}^{k}$, there exists a graph $G$ of $n$ nodes so that $\E_{\vy\sim \mathcal{M}^{k=1}(G)}[I_G(\vy)]\le \opt-\left(\frac{n}{4} - (e^{\epsilon}+1)\right)$ where $\opt = \max_{S:|S|=1} I_G(S)$. 

Moreover, for any $2\le d$, there exists a graph $G$ of $n$ nodes and maximum degree upper bounded by $d$ so that $\E_{\vy\sim \mathcal{M}^{k=1}(G)}[I_G(\vy)] \le (1-1/e) \opt-\frac{1}{10e^{\epsilon}}
\Bigl(1-\frac{d}{d-1}
\Bigl(\frac{1}{e}+\frac{1}{10}\Bigr)\Bigr)n
+\Bigl(1
-\frac{d}{d-1} \Bigl(\frac{1}{e}+\frac{1}{10}\Bigr)\Bigr)\Bigl(1+\frac{1}{10e^{\epsilon}}\Bigr)$.
\end{proposition}
%}

When the maximum degree of $G$ is $n-1$, we use the \emph{star graph}, since it differs from a totally isolated graph by the connection of a single node; a node-DP algorithm must output similar seed sets on these two graphs and thus yields low-quality seeds. When the maximum degree satisfies $2 \le d < n-1$, we construct $G'=(V,p')$ on $n$ nodes consisting of $l$ disjoint $d-1$-regular trees $S_1,\dots,S_l$, each with roughly $(n-1)/l$ nodes and edges $p'_e=1$ within trees and $p'_e=0$ otherwise, and obtain $G=(V,p)$ by adding one node $v$ that connects the root nodes of two trees $S_i$ and $S_j$. Under node DP, the algorithm must produce outputs with similar probabilities on $G$ and $G'$, so it cannot assign substantially higher probability to the high-quality seeds in $G$ than in $G'$, where such seeds are ineffective, leading to low expected influence. Based on these constructions, the following proposition (proved in \Cref{proof:prop:edge:dp:neg}) further shows that no edge-DP seeding algorithm can output high-quality seeds either, and both hardness results extend to the approximate $(\epsilon,\delta)$-DP setting where \Cref{eq:DP-def} allows an additive $\delta$; see \Cref{proof:prop:edge:dp:neg}.

%\fang{This statement is problematic.  In particular, you should have additional condition on the maximum degree in the second case($d<n-1$).  Otherwise, if a graph has no edge, all algorithms have the same value.  The wording is also not clear: "If the maximum degree is $n-1$, there exists a graph $G$"  You should first have a graph before defining maximum degree.{\color{blue}solved}}

%\fang{It may be better if this result can be translated to the following form: for any $\epsilon$, there exists $\alpha_\epsilon$ so that all $\epsilon$-dp algorithm cannot have additive error better than $\alpha_\epsilon$, $\E[I_{G}(\vy)]<(1-1/e)\opt-\alpha_\epsilon n$.} 

{%\color{red}
\vspace{-10pt}
\begin{proposition}[Hardness of Influence Maximization under Edge DP]\label{prop:edge:dp:neg}
For any $n>1$, $\epsilon>0$, and $\epsilon$-edge DP seeding algorithm $\mathcal{M}^{k}$, there exists a graph $G$ on $n$ nodes whose maximum degree satisfies $d \ge 2$ such that $ \E_{\vy\sim \mathcal{M}^{k=1}(G)}[I_G(\vy)] \le (1-1/e)\opt -\frac{1}{10e^{\epsilon}}
\Bigl(\frac{4}{5}-\frac{2}{e}\Bigr)n
+\frac{4}{5}
-\frac{2}{e}$ where $\opt = \max_{S:|S|=1} I_G(S)$.
\end{proposition}}

%\fang{should be "there exists a graph $G$ on $n$ nodes and $\alpha_\epsilon \ge \frac{1}{1000e^{\epsilon}}$, so that $(1-1/e)\max_{S:|S|=1} I_G(S) - \E_{\vy\sim \mathcal{M}^{k=1}(G)}[I_G(\vy)] \ge \alpha_\epsilon n$  The moreover part has the same issue.{\color{blue}solved}}

%\mar{Would it be important to show if this hardness result is still unavoidable when $\delta>0$ in $(\epsilon,\delta)$-DP?}

Compared to \Cref{thm:original}, the above results show that node-DP (or edge-DP) algorithms cannot perform close to $(1-1/e)\opt$ even with full knowledge of $G$. In \Cref{proof:prop:edge:dp:neg} (\Cref{prop:epsilon-delta-hardness}), we show that similar lower bounds hold even if we consider additive relaxations of the node and edge DP, allowing the $\Pr[\mathcal{M}^k(G) = \vy]\le e^\epsilon \Pr[\mathcal{M}^k(G') = \vy]$ condition to be violated up to a $\delta$ additive factor. Building on the above hardness results for node- and edge-level DP, we next design ISDP algorithms that achieve performance comparable to \Cref{alg:original}.

\section{Differentially Private Seeding Algorithms} \label{sec:mechanisms}
In this section, we provide two DP seeding algorithms that satisfy the DP requirements either centrally or locally. First in \Cref{sec:exp}, we use an exponential mechanism that ensures that the seed selection output is centrally DP. Next in \Cref{sec:rr}, we use the randomized response algorithm to achieve local DP according to the criteria set forth in \Cref{def:ISDP}. \Cref{thm:exp:acc,thm:rr:acc} provide our sample complexity bounds in each case to achieve $\epsilon$-ISDP with a $(1-1/e)\opt - \alpha n$ performance guarantee. In particular, we observe that the sample complexity bound for achieving local DP is much harsher at $O(k^3\epsilon^{-2k^2} \ln n /\alpha^2)$ compared to $O(k \ln n/(\alpha\epsilon))$ for central DP. This exponential separation of the dependency of sample complexity on $k$ is reflected in our simulation results in Section \ref{sec:sim}, where the exponential mechanism shows a faster improvement in performance with an increasing number of influence samples and its output has a significantly higher expected spread size when the privacy budget ($\epsilon$) is low.  
\subsection{Central DP with Exponential Mechanism}\label{sec:exp}

% \section{Differentially private inf-sample algorithm}
In this section, we design a differentially private (DP) seeding algorithm based on the exponential mechanism, building upon and modifying \Cref{alg:original}. To ensure privacy protection, we replace the seed selection step in \Cref{alg:original_single} with an exponential mechanism that selects a seed node according to the softmax distribution induced by the scores under differential privacy. Formally, to achieve $\epsilon$-DP, let the output set be $\mathcal{V}$ and the scoring function be $u:\{0,1\}^{n\times m}\times\mathcal{V}\to\mathbb{R}$ with sensitivity $\Delta(u):=\max_{v\in\mathcal{V},\,\vx{\sim}\vx'}|u(\vx,v)-u(\vx',v)|$; the exponential mechanism $\mathcal{M}_{exp}^{\epsilon}(\vx,S)$ selects $v\in\mathcal{V}$ with probability proportional to $\exp\!\big(\epsilon\,u(\vx,v)/(2\Delta(u))\big)$ \citep{dwork2014algorithmic}. In our setting, $\mathcal{V}=[n]\setminus S$ and $u(\vx,v)=I_\vx(v\mid S)=\tfrac{n}{m}\,\big|C_\vx(S\cup\{v\})\setminus C_\vx(S)\big|$, where $C_\vx(S)$ denotes the set of influence samples whose activated nodes are reachable from the seed set $S$; the sensitivity is $\Delta(u)=\tfrac{n}{m}$. Given $\epsilon>0$ and an integer $k$, we define the DP seeding algorithm $\mathcal{M}^{k,\epsilon}_{\mathrm{exp}}(\vx)$, which runs \Cref{alg:original} but replaces each single-seed selection step of \Cref{alg:original_single} with its DP variant $\mathcal{M}^{\epsilon}_{\mathrm{exp}}(\vx,S)$ in \Cref{alg:dp_single}, i.e., an exponential-mechanism selection using $I_\vx(\cdot\mid S)$ as the score.

{\small

\begin{algorithm}
\caption{$\epsilon$-ISDP {exponential mechanism}, $\mathcal{M}_{exp}^\epsilon(\vx,S)$}\label{alg:dp_single}
  \begin{algorithmic}[1]
    \Require $m$ influence samples $\vx = (x^1, \dots, x^m)^\top$, a seed set $S$, and privacy budget $\epsilon \geq 0$.
    \For{$v\in [n]\setminus S$}
    \State Compute $I_\vx(v|S)$
    \EndFor
    \State Return $v$ with probability proportional to $\exp\left(\frac{\epsilon m}{2n} I_\vx(v|S)\right)$ \label{iter:soft-max}
  \end{algorithmic}
\end{algorithm}}

% \fang{Complexity is $O(nmk)$.}\ar{Thanks, fixed throughout!}

% \begin{algorithm}
% \caption{$\epsilon$-differentially private inf-sample algorithm, $\mathcal{M}_{exp}$}
%   \begin{algorithmic}[1]
%     \Require $m$ influence samples $\vx = A_1, \dots, A_m$ where $A_j \in \{0,1\}^{n}$ for all $j\in [m]$ and $\epsilon_1, \dots, \epsilon_k$ with $\sum_i \epsilon_i = \epsilon$.
%     \State Set $\Lambda^* = \emptyset$.
%     \For{$i\in [k]$}
%     \State Update $\Lambda^* = \Lambda^*\cup \mathcal{M}_i(\vx,\epsilon_i,\Lambda^*)$
%     \EndFor
%   \end{algorithmic}
%   Return $\Lambda^*$
% \end{algorithm}

Note that a naive algorithm that runs the exponential mechanism directly on all $k$ seeds at once is not efficient because the size of the output space is combinatorially large $\Omega(n^k)$. To ensure that our algorithm is efficient, we sequentially run the exponential mechanism $k$ times to estimate and choose seeds similar to the greedy algorithm of \citet{costlyseeding}, but reducing their time complexity to $O(nmk)$. \Cref{thm:exp:inf} shows that our modified greedy algorithm ($\mathcal{M}_{exp}^{k,\epsilon}(\vx)$) that applies the exponential mechanism ($\mathcal{M}_{exp}^\epsilon(\vx,S)$) for single seed selection $k$ times, is $\epsilon$-ISDP following \Cref{def:ISDP}, and also provides a tail bound on the approximation error for maximization of $I_\vx$ using $\mathcal{M}_{exp}^{k,\epsilon}(\vx)$. Its proof in \Cref{proof:thm:exp:inf} follows from the facts that $I_\vx$ is a submodular function, and the exponential mechanism in \Cref{alg:dp_single} can be seen as a noisy maximum function (softmax) that implements an approximate greedy step in \Cref{thm:exp:utility}.  

\vspace{-0.5em}
\begin{theorem}[label = thm:exp:inf]
Given $k \leq n \in \mathbb{N}$, $\epsilon>0$ and $m\in \mathbb{N}$ influence samples, $\vx$, $\mathcal{M}^{k, \epsilon}_{exp}(\vx)$ is $\epsilon$-ISDP and outputs a set of $k$ nodes which, for all $t>0$, satisfies:
\begin{align*}
&\Pr\left[I_\vx(\mathcal{M}^{k, \epsilon}_{exp}(\vx))\le (1-e^{-1})\max_{S:|S| = k}I_\vx(S)- \frac{2k^2n}{\epsilon m}(\ln n+t)\right]  \le ke^{-t},
\end{align*}
where the probability is over the randomness of the algorithm run for fixed $\vx$.
\end{theorem}
\vspace{-0.5em}
{This theorem shows that the exponential mechanism $\mathcal{M}^{k,\epsilon}_{\mathrm{exp}}(\vx)$ achieves $\epsilon$-ISDP while maintaining strong utility guarantees. Specifically, with high probability, the influence achieved by the private seed set is close to the optimal non-private value on the same influence samples $\vx$. The term $(1-1/e)$ reflects the classical approximation ratio for submodular maximization, while the additive term $\tfrac{2k^2 n}{\epsilon m}(\ln n + t)$ quantifies the accuracy loss due to privacy and limited samples. In particular, the bound reveals that utility improves as the number of influence samples $m$ increases or the privacy budget $\epsilon$ becomes larger.} Equipped with \Cref{prop:inf_ch}, we can derive the following error bound with respect to the optimal value $I_G$ of the influence maximization problem (proved in \Cref{proof:thm:exp:acc}):  

\vspace{-0.5em}
\begin{theorem}[label = thm:exp:acc]
Given graph $G$ on $n\ge 2$ nodes and $k\le n$, for any $0<\alpha\le 1$ and $\epsilon>0$, $\mathcal{M}^{k,\epsilon}_{exp}(\cdot)$ that runs \Cref{alg:original} but replaces the single seed selection steps of \Cref{alg:original_single} by its exponential mechanism variant in \Cref{alg:dp_single}, $\mathcal{M}_{exp}^{\epsilon/k}(\vx,S)$, is $\epsilon$-ISDP and its output infects at least $(1-1/e)\opt-\alpha n$ nodes with high probability, using at least $m =  \max\{\frac{12}{\alpha\epsilon},\frac{9}{\alpha^2}\}k\ln n$ influence samples in $O(knm)$ run time.
\end{theorem}
\vspace{-0.5em}

Note that if we are not concerned with privacy and set $\epsilon\gg \alpha$, then the number of samples to achieve the additive error $\alpha n$ is of the order $\alpha^{-2}k\ln n$, which improves the $k^2(\ln n/\alpha)/\alpha^3$ bound of \citet[Theorem 2]{costlyseeding}. This is because in their Algorithm 1 and Theorem 2, \citet{costlyseeding} collect $\rho=81k log(6nk/\epsilon)/\epsilon^3$ new influence samples for selecting each seed for a total of $k \left\lceil \normalcolor 81 k \log( {6nk}/{\epsilon})/{\epsilon^3} \right\rceil \normalcolor \in {O}(k^2\log({n}/{\epsilon})/{\epsilon^3})$ influence samples, whereas we observe that we can work with the same set of $m =  \max\{\frac{12}{\alpha\epsilon},\frac{9}{\alpha^2}\}k\ln n$ influence samples to select all the $k$ seeds. Moreover, our result does not violate Theorem 3 of \citet{costlyseeding} which shows that $\Omega(k^2)$ influence samples are required to achieve the $(1-1/e)\opt-\alpha n$ performance guarantee for all $\alpha<(1-1/e)/k$, because our algorithm also needs $\Omega(k^2)$ samples when $\alpha <1/k$. 

% \fy{privacy concern}
% Given $\epsilon>0$, we set $\epsilon_1, \dots, \epsilon_k$ so that $\sum_i \epsilon_i = \epsilon$, and iteratively call \Cref{alg:dp_single}.  Then by composition theorem\fy{be more careful}, the algorithm is $\epsilon$-differentially private.

% Each node can decide reporting all contagions or not.  Then the error guarantee is  
% $$O\left(\frac{1}{\min \epsilon_i}km\ln n\right) = O\left(\frac{1}{\min \epsilon_i}k^3(\ln n)^2\right)$$ when $m = k^2\ln n$

% As we have more samples, we we need to add larger noisy in the exponential mechanism, but at the same time we can have a better estimation error for non-private inf-sample algorithm.  

% Informally, if we use $m$ samples, the error between exponential mechanism and the non-private inf-sample algorithm is $O(\frac{1}{\epsilon_{DP}}km\ln n)$.  On the other hand, the error of inf-sample algorithm is $O(\frac{nk^{2/3}}{m^{1/3}})$.  Therefore, the optimal number of sample is $m = \Theta(\sqrt[4]{n^3/(k(\ln n)^3)})$ with additive error $\Theta(\sqrt[4]{k^3n^3\ln n})$.

\subsection{Local DP with Randomized Response}\label{sec:rr}

%{\color{red}
In the \emph{central} setting, privacy protection is applied by the curator after collecting all influence samples, ensuring that the published algorithm outputs do not reveal any individual’s participation. In contrast, in the \emph{local} setting, privacy is enforced during data collection itself. In contact-tracing applications, for example, it may be known that a confirmed positive case attended a particular event, but other participants may not wish to disclose their own attendance with certainty. Since individuals may not fully trust the data collector, they can protect their privacy using a \emph{randomized response} mechanism—similar to those in prior local DP studies—which allows them to probabilistically deny participation. Concretely, we achieve local $\epsilon$-ISDP by perturbing the input data via randomized response: each entry of $\vx$ is independently flipped with probability $\rho_{\epsilon}=1/(1+e^{\epsilon})$ to produce the noisy input $\tilde{\vx}$, as shown in Line~\ref{iter:local-dp-noise} of \Cref{alg:dp_rrim}.%} 
However, the added input noise will bias our estimate of each candidate seed's quality; in particular, $\E_{\tilde{\vx}}[I_{\tilde{\vx}}(S|T)]$ is not necessarily equal to $I_G(S|T)$ for a pair of subsets $S,T \subset V$ and we no longer have the benefit of applying \Cref{prop:inf_ch} directly. Rather in \Cref{alg:offset_perturb} we propose a post-processing of the noisy input data $\tilde{\vx}$ to derive an unbiased estimate of a seed set's quality. The importance of proper post-processing steps in the analysis of DP data is highlighted in other applications, e.g., in the release of census DP data, where ignoring them can lead to biased decision-making and suboptimal results \citep{ijcai2022p559}. The added variability of the estimators due to the DP noise can also degrade our estimation performance (even if not biased), which is a second concern in local DP algorithm design and opens a venue to explore alternative post-processing methods with potentially superior bias-variance tradeoff. In the remainder of this section, we first introduce \Cref{alg:dp_rrim,alg:offset_perturb} and then offer \Cref{thm:rr:acc} for their privacy and utility guarantees, addressing both concerns about bias and variance of decision making with local DP influence samples.

\vspace{-10pt}
{\small 
\begin{algorithm}[ht]
\caption{Local $\epsilon$-ISDP seeding algorithm, $\mathcal{M}_{loc}^{k,\epsilon}(\vx)$}\label{alg:dp_rrim}
  \begin{algorithmic}[1]
    \Require $m$ influence samples $\vx$, total number of seeds $k$, and privacy budget $\epsilon>0$.
    \State Set $S_0 = \emptyset$.
    \State Flip any bits of $\vx$ with probability $\rho_\epsilon = \frac{1}{e^\epsilon+1}$ and get $\tilde{\vx}$. \label{iter:local-dp-noise}  
    \For{$i\in [k]$}
    \State Find $v_i\in \argmax_v J_m(S_{i-1}\cup \{v\})$ where $J_m(S)$ is defined in \Cref{alg:offset_perturb}.\label{iter:1}
    \State $S_{i} = S_{i-1} \cup v_i$.  
    \EndFor
    \State Return $S_k$
  \end{algorithmic}
\end{algorithm}}
\vspace{-15pt}

Given a candidate seed set \(S\subseteq V\) with \(|S|=l\), an influence sample
\(x\in\{0,1\}^n\), and its perturbed version \(\tilde x\), define
\(z_S:=|S\cap x|\) and \(\tilde z_S:=|S\cap \tilde x|\). The event
\(z_S>0\) means that at least one node in \(S\) appears in the influence sample,
or equivalently, that \(S\) reaches the sampled target in that sample. Therefore,
by \Cref{prop:inf_ch}, \(I_G(S)=n\Pr[z_S>0]\). In the local privacy regime,
however, the algorithm observes only the perturbed samples \(\tilde x\), and
hence only empirical information about \(\tilde z_S\). The goal of the
post-processing step is to recover \(\Pr[z_S>0]\), or equivalently
\(\Pr[z_S=0]\), from observations of \(\tilde z_S\).

The key observation is that randomized response induces a known transition
matrix between the distribution of \(z_S\) and the distribution of
\(\tilde z_S\). Let \(f_b=\Pr[z_S=b]\) and
\(\tilde f_a=\Pr[\tilde z_S=a]\), for \(a,b=0,\ldots,l\). Conditional on
\(z_S=b\), exactly \(b\) entries among the \(l\) nodes in \(S\) are equal to one
before perturbation, and \(l-b\) entries are equal to zero. After randomized
response, the number of ones that remain one is distributed as
\(\mathrm{Bin}(b,1-\rho_\epsilon)\), and the number of zeros that flip to one is
distributed as \(\mathrm{Bin}(l-b,\rho_\epsilon)\). These two binomial random
variables are independent. Hence, for \(a,b=0,\ldots,l\),
\[
    C_{\rho_\epsilon,l}(a,b)
    :=
    \Pr[\tilde z_S=a\mid z_S=b]
    =
    \Pr\!\left[
    \mathrm{Bin}(b,1-\rho_\epsilon)
    +
    \mathrm{Bin}(l-b,\rho_\epsilon)
    =
    a
    \right].
\]
Equivalently, if \(C_{\rho_\epsilon,l}\in\mathbb R^{(l+1)\times(l+1)}\) denotes
the matrix with entries \(C_{\rho_\epsilon,l}(a,b)\), then
\[
    (\tilde f_0,\ldots,\tilde f_l)^\top
    =
    C_{\rho_\epsilon,l}(f_0,\ldots,f_l)^\top .
\]

We can also write each entry explicitly. Suppose \(z_S=b\). To obtain
\(\tilde z_S=a\), let \(r\) be the number of original one-entries that remain
one after perturbation. Then \(a-r\) of the original zero-entries must flip to
one. Therefore,
\[
    C_{\rho_\epsilon,l}(a,b)
    =
    \sum_{r=\max\{0,a-(l-b)\}}^{\min\{a,b\}}
    \binom{b}{r}(1-\rho_\epsilon)^r\rho_\epsilon^{b-r}
    \binom{l-b}{a-r}\rho_\epsilon^{a-r}
    (1-\rho_\epsilon)^{l-b-a+r}.
\]
This expression simply enumerates all ways to obtain \(a\) observed ones after
randomized response: \(r\) retained ones and \(a-r\) flipped zeros.

If \(C_{\rho_\epsilon,l}\) is invertible, we can undo the effect of randomized
response at the distributional level. Specifically, from the empirical
distribution \(\tilde{\vf}\) of \(\tilde z_S\), we solve
\(\tilde{\vf}=C_{\rho_\epsilon,l}\vf\) to estimate the distribution \(\vf\) of
\(z_S\). Since the empirical distribution satisfies
\(\mathbb E[\tilde{\vf}]=C_{\rho_\epsilon,l}\vf^\star\), where \(\vf^\star\)
denotes the true distribution of \(z_S\), this post-processing gives
\(\mathbb E[\vf]=\vf^\star\). Hence, \(1-f_0\) is an unbiased estimator of
\(\Pr[z_S>0]\), and \(J_m(S)=n(1-f_0)\) is an unbiased estimator of \(I_G(S)\).
\Cref{alg:offset_perturb} summarizes the procedure.

{\small \begin{algorithm}
\caption{Unbiased estimate for the size of the reachable set $J_m(S)$}\label{alg:offset_perturb}
  \begin{algorithmic}[1]
    \Require $m$ perturbed influence samples $\tilde{\vx} = (\tilde{x}^{1},\ldots,\tilde{x}^{m})$,  their perturbation probability $\rho_\epsilon$, and a candidate set $S\subset V$ with size $|S| = l$.
    \State Compute $\tilde{\vf}=(\tilde{f}_0, \tilde{f}_1, \dots, \tilde{f}_l)\in \mathbb{R}^{l+1}$ the empirical distribution of $\tilde{z}_S =|S\cap \tilde{x}|$ so that $\tilde{f}_a = \frac{1}{m}\sum_{t=1}^{m} \mathbf{1}[|S\cap \tilde{x}^t| = a]$ for all $a=0,1,\dots, l$. \label{iter:empirical-dist}
    \State Solve for $\vf = (f_0, f_1, \dots, f_l)\in \R^{l+1}$ so that $\tilde{\vf} = C_{\rho_\epsilon,l}\vf$. \label{iter:postprocessing}
    % $f = \argmin_{f'\in \Delta_{l+1}} \|A_{\rho_\epsilon,l}f'-\tilde{f}\|_2$ where $\Delta_{l+1}$ is the collection of probability on set ${0,1,\dots, l}$. 
    \State Return $J_m(S) = n(1-f_0)$ 
    \label{iter:postprocessing-return}
  \end{algorithmic}
\end{algorithm}}
%\vspace{-10pt}
%\vspace{-10pt}

\begin{proposition}[label = prop:feasible, name = Feasibility of Unbiased Post-processing]
Given any positive integer $l\ge 1$ and $\rho_\epsilon\in [0, 1/2)$, the matrix $C_{\rho_\epsilon,l}$ is invertible.
\end{proposition}

Note that when $\rho_\epsilon = 1/2$, $\tilde{\vx}$ consists of uniform random bits and we cannot recover $\Pr[z_S > 0]$ from perturbed influence samples. \Cref{prop:feasible} shows that we can recover $\Pr[z_S > 0]$ for any $\rho_\epsilon < 1/2$.  

% The above algorithm is the greedy algorithm on function $\tilde{J}(S) = \frac{1}{m}\sum_{t = 1}^m \bm{1}[S\cap \tilde{x}^t\neq \emptyset]$.  Let $\alpha_k = \frac{n}{(1-\rho_\epsilon)^{k}}$ and $\beta_k = n-\alpha_k = n(1-1/(1-\rho_\epsilon)^{k})$ for all $k\in \mathbb{N}$.  The lemma below show that $\alpha_{|S|}\tilde{J}(S)$ is close to influence function of $S$, $I_G(S)$.  

% \begin{lemma}
% For each subset of nodes $S\subseteq V$, 
% $$0\le \Pr[S\cap \tilde{x}\neq \emptyset]-(1-\rho_\epsilon)^{|S|}\frac{1}{n}\E_G[I(S)]\le (1-(1-\rho_\epsilon)^{|S|}).$$  
% Moreover, 
% $$\Pr\left[\left|\frac{1}{m}\sum_{t = 1}^m \bm{1}[S\cap \tilde{x}^t\neq \emptyset]-\Pr[S\cap \tilde{x}\neq \emptyset]\right|\ge \delta\right]\le \exp(-2m\delta^2)$$
% \end{lemma}
\vspace{-0.5em}
\begin{theorem}[label = thm:rr]
Given $\epsilon>0$, $m$ influence samples on a graph with $n$ nodes and integer $k<n$, \Cref{alg:dp_rrim} is locally  $\epsilon$-ISDP, and its output satisfies: $I_G(\mathcal{M}_{loc}^{k,\epsilon}(\vx))>(1-1/e)\opt-\delta,$ with probability $1-2n^k\exp\left(-{m\delta^2}/{(2k^2n^2V_{\rho_\epsilon, k}^2)}\right),$ where $V_{\rho_\epsilon, k} = O((1/2-\rho_\epsilon)^{-k^2})$.
\end{theorem}
\vspace{-0.5em}
The above theorem, proved in \Cref{proof:thm:rr}, provides a tail bound on the approximation error of $\mathcal{M}_{loc}^{k,\epsilon}$ given a fixed number of influence samples. We can convert the above result to a high probability error bound by choosing $m = \Omega({k^3V_{\rho_\epsilon,k}^2\ln n}/{\alpha^2})$ which for $k$ constant and $\epsilon \gg \alpha$ improves the sample complexity ($1/\alpha^3$) in \Cref{thm:original} (complete proof in \Cref{proof:thm:rr:acc}):

\vspace{-0.5em}
\begin{theorem}[label = thm:rr:acc]
Given graph $G$ with $n\ge 2$ nodes and $k\le n$, for any $0<\alpha\le 1$ and $\epsilon>0$, $\mathcal{M}_{loc}^{k,\epsilon}$ is locally $\epsilon$-ISDP and its output infects at least $(1-1/e)\opt-\alpha n$ nodes with high probability, using at least $m = O({k^3V_{\rho_\epsilon,k}^2\ln n}/{\alpha^2})$ influence samples in $O(nk^4 + k n^2 m)$ run time, where $V_{\rho_\epsilon, k} = O((1/2-\rho_\epsilon)^{-k^2})$.
\end{theorem}
\vspace{-0.5em}

We conclude this section by considering two scenarios: the non-private regime $\epsilon\to \infty$ and the full-privacy regime $\epsilon\to 0$. First, in the non-private regime ($\epsilon\to \infty$), $\rho_\epsilon\to 0$, which causes the matrix $C_{\rho_\epsilon,l}$ and its inverse to converge to the identity matrix, so we can take $V_{\rho_\epsilon, k}$ to be a constant arbitrarily close to one. As a result, $m = O\left(\frac{k^3\ln n}{\alpha^2}\right)$ samples are sufficient, which is comparable to \cref{thm:original} with an additional factor of $k$, but improves the dependence on $\alpha$. We may reduce the dependence on $k$ from $k^3$ to a smaller order, using techniques in adaptive data analysis~\cite{bassily2016algorithmic}. Informally, our analysis requires that the query error in any possible seed set $S$ with $|S|\le k$ is small enough (\cref{lem:concentrate_infl}) so that the output of the greedy algorithm is good enough. However, we may not need such stringent conditions, because the greedy algorithm does not necessarily query all possible seed sets. Alternatively, adaptive data analysis provides a procedure to support adaptive queries from the greedy algorithm with a better error bound. \footnote{Interestingly, the procedure in \cite{bassily2016algorithmic} adds Gaussian noise to each query and achieves a small error guarantee using techniques from an exponential mechanism similar to ~\cref{thm:exp:acc}.} Furthermore, in the non-private regime ($\epsilon\to\infty$), our exponential mechanism and the local DP seeding algorithm are identical, allowing us to use the analysis in
\cref{thm:exp:inf} and derive an $O\left(\frac{k\ln n}{\alpha^2}\right)$ upper bound for the required number of influence samples.
On the other hand, if $\epsilon\to 0$, $V_{\rho_\epsilon, k} = O((1/2-\rho_\epsilon)^{-k^2})$ in \Cref{thm:rr,thm:rr:acc} can be simplified as follows: $V_{\rho_\epsilon, k} = O((1/2-\rho_\epsilon)^{-k^2}) = O((1/2-1/(e^\epsilon+1))^{-k^2}) = O([2(e^\epsilon+1)/(e^\epsilon-1)]^{k^2}) =O(\epsilon^{-k^2})$.  %\fang{the regularization idea we discussed is for $\epsilon\to 0$ case ($C_{\rho_\epsilon,l}$ becomes singular) and not necessary for no privacy setting.}

\section{Simulations with Empirically-Grounded Network Data Sets}\label{sec:sim}% and the Bias-Variance Dilemma of Post-Processing
We begin by evaluating our proposed framework for data collection and intervention design using highly sensitive data on sexual activity in an HIV prevention context in \Cref{sec:msm}, while additional sanity checks based on synthetic data generated from Erdős–Rényi networks are provided in Appendix~\ref{sec:app:synthetic}.

\subsection{Simulations on Empirically-Grounded Sexual Activity Networks}\label{sec:msm} We construct a dynamic network of men who have sex with men (MSM) using a separable temporal exponential random graph model (STERGM), estimated from the ARTnet study \citep{weiss2020ARTnet}, a cross-sectional egocentric survey conducted in the United States (2017–2019) with 4,909 participants reporting 16,198 sexual partnerships. Our analysis focuses on San Francisco due to data availability and its representativeness for the MSM population. The STERGM estimation procedure is summarized in Appendix~\ref{sec:app:stergm-estimation}.

Using the estimated STERGM, we simulate sexual partnerships among 1,000 individuals over 120 weekly periods, with fixed node attributes (e.g., race, age). We aggregate every 12 weeks as one quarter. In each quarter, we randomly select 150 initial nodes and identify all individuals reachable through sexual relationships, yielding 150 influence samples per quarter and 1,500 samples in total. A second, independently generated set of influence samples is used to evaluate the seed sets output by our algorithm.

Unlike settings that assume a latent network and independent‐cascade dynamics, we directly construct influence samples to mimic contact tracing: each cascade begins with an index case and recursively includes their sexual partners. This approach reflects realistic exposure pathways without requiring full knowledge of the underlying network. To protect the privacy of individuals involved in these sensitive behavioral data, we implement both central and local privacy mechanisms: the curator adds noise to released outputs, while individuals may randomize their reported connections using randomized response.

We ran our privacy-preserving algorithm on influence samples constructed from ARTnet study data. In this case, the chosen seeds are high-risk nodes whose protection provides the most overall benefit in terms of preventing large disease cascades. In the simulation setup of \Cref{fig:HIV-network-k-1020-exp-rr}, we consider $k=18$ and $k=20$ seeds, the number of influence samples ranges from zero to $1500$, $m \in \{0, 500, 1000, 1500\}$. The case $m=0$ corresponds to the absence of input information where the choice of 18 or 20 seeds will be uniformly random. With no input information at $m=0$, there are no privacy risks to consider. 

\begin{figure}[htb]
\centering
\includegraphics[width=0.6\textwidth]{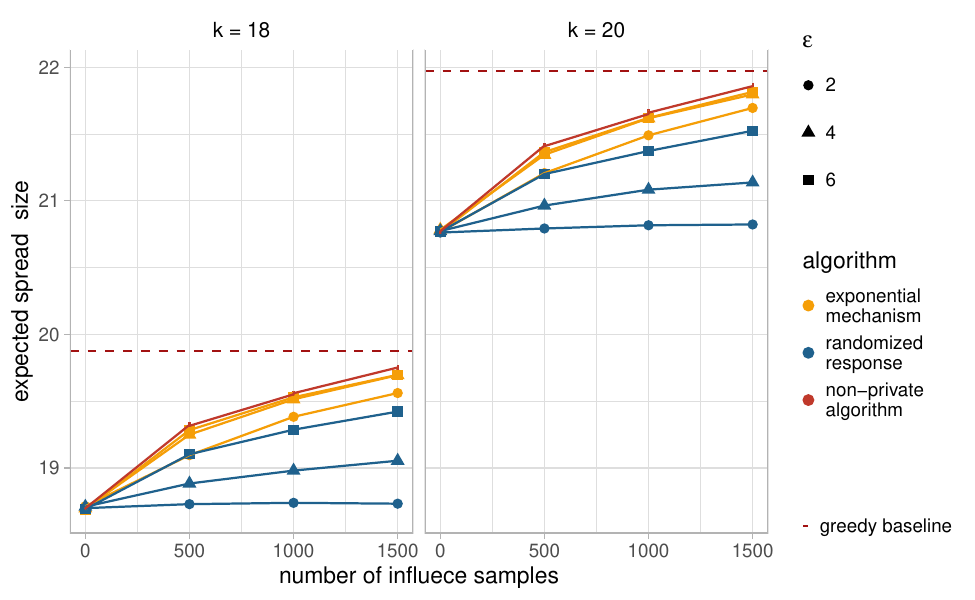}
\caption{\small 
Performance of our privacy-preserving seeding algorithms—the exponential-mechanism and randomized-response methods—on the MSM dataset. 
We report the expected influence $I_G(\cdot)$ across privacy budgets ($\epsilon$) as the number of influence samples ($m$) increases, and compare against two non-private baselines: influence-sample seeding and the deterministic greedy algorithm with complete network information. 
Each point represents the mean over 1,000 trials obtained from 50 independent collections of $m$ influence samples, each evaluated 20 times. 
Error bars show $95\%$ confidence intervals (mostly too small to be visible). 
The expected spread size for each seed set is estimated using an additional 1,000 influence samples.}
\label{fig:HIV-network-k-1020-exp-rr}
\end{figure}

In the \cref{fig:HIV-network-k-1020-exp-rr}, we see that the exponential mechanism outperforms the randomized response mechanism as the latter provides stronger (local) privacy protections. We also observe that low values of privacy budget $\epsilon$ lead to earlier saturation in lower expected spread sizes. In those cases, our maximum number of influence samples is not enough to counteract the random noise introduced by privacy mechanisms, so the value of network information diminishes with the decreasing privacy budge -- cf. \citet[Section 4]{costlyseeding}. Increasing the privacy budget alleviates this effect, as we see a faster increase in the expected spread size with increasing $m$ at higher values of $\epsilon$. In fact, the performance of both mechanisms approaches the original (``non-private") influence sample greedy seeding algorithm (\cref{alg:original}), as $\epsilon \to\infty$. For example, in the randomized response mechanism (\Cref{alg:dp_rrim}), as the privacy budget increases, the flip probability of the entries in the influence samples $x \in \{0,1\}^{n\times m}$ (\Cref{alg:dp_rrim}, line~\ref{iter:local-dp-noise}) decreases and the matrix $C_{\rho_\epsilon,l}$ in \cref{eq:c_rho} approaches the identity matrix, so the performance is similar to greedy non-private seeding on influence samples in \Cref{alg:original}. For the exponential mechanism of \Cref{sec:exp}, at each step the mechanism chooses the seeds probabilistically with probabilities proportional to the exponential of their marginal contribution to the spread size computed on the influence samples. In this case, increasing the privacy budget ($\epsilon\to\infty$) increases the probability of choosing the seed with the largest marginal contribution. \Cref{fig:HIV-network-k-1020-exp-rr} also shows the deterministic baseline for greedy with complete network information to be compared with the exponential mechanism and randomized response algorithms at different $\epsilon$ values with increasing $m$. As we gain better information about the network by increasing the number of influence samples $(m\to\infty)$ and the privacy budget $(\epsilon\to\infty)$, the performance of both algorithms will converge to the deterministic greedy algorithm with complete network information.

\section{Discussions, 
 managerial insights and future directions}\label{sec:conc}

%In this paper, we propose a new notion of differential privacy using adjacency of influence samples. This notion may be extended to various problems that have an underlying network structure. Examples include viral content and rumor cascades on social networks, polls in elections from voters who influence each other's opinions, and medical history of past disease contagions on epidemiological networks.  Note that the input of an influence sample DP algorithm is different from the inputs of node- or edge-DP algorithms -- capturing different notions of privacy loss that are not formally comparable; specifically, in influence sample DP an individual may decide whether or not to report their particular value for an influence sample but they cannot change the data generation process. Sometimes, the underlying network may only exist as an abstract construct that is a consequence of other physically well-defined relations (e.g., the sexual activity network in \Cref{sec:sim}). In such cases, the network cascade data are the only available input data that can receive meaningful privacy protection. Our results address \emph{entry-level} privacy by defining adjacency between cascade entries in the input data set. One can generalize this to \emph{individual-level} privacy by considering two adjacent inputs if they differ in one column (an individual can decide whether or not to report their entire contagion history).
\paragraph{Comparison between node, edge, and influence-sample DP.}
We introduce a new notion of differential privacy—\emph{influence-sample differential privacy} (ISDP)—which defines adjacency based on differences in observed diffusion samples rather than changes to the underlying network structure. This formulation is particularly suited to applications where the network is unknown or cannot be reliably reconstructed due to privacy constraints, such as HIV prevention, election polling, or information diffusion on social media. In contrast to node- or edge-level DP, ISDP does not assume access to the full network. When the network is latent, modifying a single node or edge can alter the entire diffusion process, thereby changing many entries in all observed cascades. Satisfying node or edge DP in such cases would require adding excessive noise, effectively destroying utility. ISDP circumvents this limitation by defining adjacency at the level of individual cascade participation—whether a person appears in a given diffusion sample—offering a meaningful privacy guarantee even without network knowledge.

This distinction highlights that different definitions of adjacency can lead to fundamentally different trade-offs between privacy and utility (\Cref{prop:node:dp:neg,prop:edge:dp:neg} versus \Cref{thm:exp:acc,thm:rr:acc}). Our goal is not to claim ISDP as superior to node or edge DP, but to emphasize that the appropriate notion of adjacency should reflect the structure and semantics of the data. In networked settings, individuals’ private data—be it census, medical, or behavioral records—are inherently correlated through network ties. Those supporting node or edge DP may argue that adjacency should depend on the network structure rather than the traditional Hamming distance between database entries \citep{rezaei2019privacy}. Conceptually, ISDP protects whether an individual is \emph{directly or indirectly connected} to high-risk individuals as inferred from observed cascades, rather than the presence of network edges themselves. By focusing on influence samples, ISDP captures the most relevant privacy risk in sensitive contexts while maintaining practical utility for data-driven interventions.

\paragraph{Efficiency and robustness.}
Our algorithms have polynomial running times in the size of the graph, with a time complexity of $O(knm)$ and $O(nk^4+kn^2m)$ for $\mathcal{M}^{k, \epsilon}_{exp}(\vx)$ and $\mathcal{M}_{loc}^{k,\epsilon}(\vx)$, respectively, for any constant privacy budget $\epsilon>0$. To improve running times, one may consider improving the iterative sampler for the exponential mechanism in \cref{alg:dp_single}, or designing a better post-processing step for Line~\ref{iter:postprocessing} in \cref{alg:offset_perturb}.
On the other hand, our ISDP randomized response algorithm $\mathcal{M}_{loc}^{k,\epsilon}(\vx)$ is automatically robust against independent random noise in influence samples. In contrast, it is unclear how well our ISDP exponential mechanism $\mathcal{M}^{k, \epsilon}_{exp}(\vx)$ performs in the presence of noise.

\paragraph{Conclusions and future directions.} By focusing on the utility of influence samples for influence maximization, our approach ensures that each user’s participation in any cascade remains private—either locally or centrally—while preserving the statistical validity of the cascades for influence estimation. We formally protect whether an individual appears in a given influence sample, granting participants plausible deniability regarding their involvement in diffusion events. Our framework applies broadly to settings such as email communication, social messaging, and ad propagation, where actions are observed over time but the underlying network is unknown or infeasible to collect in full. In future work, we plan to extend ISDP to temporally evolving data, investigating how sequential data collection affects both privacy guarantees and intervention effectiveness. Such extensions may lead to new, consequentialist notions of privacy that remain robust under varying sampling distributions and randomization noise \citep{kifer2014pufferfish,liu2020learning}.

{\small
\section*{\small{Code and Data Availability}}
The MSM sexual behavior cascade data are derived from the ARTnet study (\url{https://github.com/EpiModel/ARTnet}). Due to confidentiality agreements, the dataset is not publicly available; access requires a memorandum of understanding (MOU) and approval from the study’s Principal Investigator, Samuel Jenness (Emory University). The analysis codes used to produce the results and figures are available at \url{https://github.com/aminrahimian/dp-inf-max}.
}

%\mar{Exponential mechanism guarantees a higher pay off than the randomize response for $k=1$}
%\cite{doi:10.1073/pnas.1510612113} vulnerable group and privacy on graph
% \cite{sealfon2016shortest} other notion of differential privacy.
% item vs user level privacy~\cite{liu2020learning}

%\mar{Stronger notions of privacy where a node or an edge may be removed from entire influence samples}

%% file: sections/appendix.tex
% \section{Proof and details for \texorpdfstring{\Cref{sec:pre}}{sec:pre} (Lower bounds of node- and edge-DP)}\label{app:proofs:sec:pre}
\section{Proofs and Supplementary Results}
{%\color{red}
\subsection{Proof of \texorpdfstring{\Cref{prop:node:dp:neg}}{prop:node:dp:neg}}\label{proof:prop:node:dp:neg}
For the graph with maximum degree $n-1$, consider the isolated graph $G' = (V,p')$ of $n$ nodes where $p'_e = 0$ for all possible pairs $e$.  Given any $\epsilon$ node-DP mechanism $\mathcal{M}$, there exists a node $v_M\in V$ such that $\Pr[\mathcal{M}^{k=1}(G') = v_M]\le 1/n,$ according to the pigeonhole principle. Then we define a star graph $G = (V,p)$ of $n$ nodes with center $v_M$ where $p_{(v_M,v)} = 1/2$ for all $v\in V$ and $v\neq v_M$ and $p_{(u,v)} = 0$ if $u,v\neq v_M$.

First, the center node can in expectation infect $(n-1)/2+1$ of the nodes in $G$, therefore, $\opt\ge I_G(\{v_M\}) = (n+1)/2$. On the other hand, because $ \mathcal{M}^k$ is $\epsilon$ node-DP and $G$ and $G'$ are node-adjacent,  $\Pr[\mathcal{M}^{k=1}(G) = \{v_M\}]\le e^\epsilon \Pr[\mathcal{M}^{k=1}(G') =  \{v_M\}]\le {e^\epsilon}/{n}$. 
Therefore, 
\begin{align*}
    \E_{\vy\sim \mathcal{M}^{k=1}(G)}[I_G(\vy)]
    = &\E[I_G(\vy)|\vy\neq  \{v_M\}]\Pr[\mathcal{M}^{k=1}(G)\neq \{v_M\}]+\E[I_G( \{v_M\})]\Pr[\mathcal{M}^{k=1}(G)=  \{v_M\}]
    \\\le& (1+1/2+(n-2)/4)+n\cdot {e^\epsilon}/{n} = 1+e^\epsilon+n/4,
\end{align*}

For graphs with maximum degree $2 \le d < n-1$, we construct $G'=(V,p')$ consisting of $l$ disjoint $(d\!-\!1)$-regular trees $S_1,\ldots,S_l$, each of size about $(n-1)/l$ (a complete $d-1$-regular tree truncated to the largest depth not exceeding $(n-1)/l$, with the last level partially filled if necessary). For each edge $e$, set $p_e'=1$ if $e$ belongs to some $S_i$ and $p_e'=0$ otherwise. By the pigeonhole principle, there exist $a$ trees $S_{i_1},\ldots,S_{i_a}$ such that $\Pr[\mathcal{M}^{k=1}(G') \subset S_{i_1}\cup\cdots\cup S_{i_a}] \le a/l$.

We then obtain $G=(V,p)$ by adding a new node $v$ and connecting it to the roots of $S_{i_1},\ldots,S_{i_a}$. Any $u \in S_{i_1}\cup\cdots\cup S_{i_a}$ now reaches all nodes in these $a$ trees, giving $I_G(u)=a(n-1)/l+1$, while any $u \notin S_{i_1}\cup\cdots\cup S_{i_a}$ reaches only its own tree, so $I_G(u)=(n-1)/l$.

% \fang{did we define $I_G$?  I recall we only have $I_\mathcal{G}$ and $I_g$.  Do we need to define node DP on weighted graph?  Here the input also consists of $p$} {\color{red}Yuxin: we change the $\mathcal{G}$ to $G$, and use $\mathcal{G}$ to denote the distribution of subgraphs randomly drawn under the IC model to make $g \sim \mathcal{G}$ is correct expression}

Because $\mathcal{M}$ is $\epsilon$-node DP and $G$ and $G'$ differ by only one node, we have
$\Pr[\mathcal{M}^{k=1}(G)\subset S_{i_1}\cup\cdots\cup S_{i_a}] \le a e^{\epsilon}/l$. Hence
\[
\E_{\vy\sim \mathcal{M}^{k=1}(G)}[I_G(\vy)]
\le \frac{a e^{\epsilon}}{l}\Bigl(\frac{a (n-1)}{l}+1\Bigr)
     +\Bigl(1-\frac{a e^{\epsilon}}{l}\Bigr)\frac{n-1}{l}
= \Bigl(\frac{1}{a}+\frac{(a-1)e^{\epsilon}}{l}\Bigr)\frac{a (n-1)}{l}
  +\frac{a e^{\epsilon}}{l}.
\]
Comparing with $\opt = a (n-1)/l$ %\fang{$1+an/l$? You added an additional node $v$}
gives
\[
(1-1/e)\opt - \E_{\vy\sim\mathcal{M}^{k=1}(G)}[I_G(\vy)]
\ge
\Bigl(\frac{a-1}{a} - \frac{1}{e} - \frac{(a-1)e^{\epsilon}}{l}\Bigr)\frac{a (n-1)}{l}
- \frac{a e^{\epsilon}}{l}.
\]

We choose $l$ so that $\frac{a-1}{a} - \frac{1}{e} - \frac{(a-1)e^{\epsilon}}{l} = \frac{1}{10}$. 
Since the coefficient of $n$ in the resulting lower bound is increasing in $a$ and $a\le d$, setting $a=d$ gives
\[
(1-1/e)\opt-\E_{\vy\sim \mathcal{M}^{k=1}(G)}[I_G(\vy)]
\ge
\frac{1}{10e^{\epsilon}}
\Bigl(1-\frac{d}{d-1}
\Bigl(\frac{1}{e}+\frac{1}{10}\Bigr)\Bigr)n
-\Bigl(1
-\frac{d}{d-1} \Bigl(\frac{1}{e}+\frac{1}{10}\Bigr)\Bigr)\Bigl(1+\frac{1}{10e^{\epsilon}}\Bigr).
\]

which completes the proof.
 \qedwhite

\subsection{Proof of \texorpdfstring{\Cref{prop:edge:dp:neg}}{prop:edge:dp:neg}}\label{proof:prop:edge:dp:neg}

Consider $G'=(V,p')$ on $n$ nodes formed by $l$ disjoint $d-1$-regular trees $S_1,\dots,S_l$, each of size about $n/l$ (with the last level truncated if needed). For each edge $e$, set $p'_e=1$ if $e$ lies in some $S_i$ and $p'_e=0$ otherwise. As in \Cref{prop:node:dp:neg}, for any $\epsilon$-edge DP mechanism $\mathcal{M}$ there exist two trees $S_i,S_j$ such that $\Pr[\mathcal{M}^{k=1}(G')\subset S_i\cup S_j]\le 2/l$.

Define $G=(V,p)$ by adding a single bridge edge $e^*$ between the roots of $S_i$ and $S_j$, with $p_{e^*}=1$ and $p_e=p'_e$ for all $e\neq e^*$. Then any $v\in S_i\cup S_j$ reaches $2n/l$ nodes in $G$, so $I_G(v)=2n/l$, whereas any $v\notin S_i\cup S_j$ reaches only its own tree, so $I_G(v)=n/l$. Because $\mathcal{M}$ is $\epsilon$-edge DP and $G$ and $G'$ differ in a single edge, we have
$\Pr[\mathcal{M}^{k=1}(G)\subset S_i\cup S_j]\le 2e^\epsilon/l$.
Hence
\[
\E_{\vy\sim \mathcal{M}^{k=1}(G)}[I_G(\vy)]
\le \frac{2e^\epsilon}{l}\cdot\frac{2n}{l}
   +\Bigl(1-\frac{2e^\epsilon}{l}\Bigr)\frac{n}{l}
= \Bigl(\frac{1}{2}+\frac{e^\epsilon}{l}\Bigr)\frac{2n}{l},
\]
and therefore
\[
(1-1/e)\opt-\E_{\vy\sim \mathcal{M}^{k=1}(G)}[I_G(\vy)]
\ge \Bigl(\frac{1}{2}-\frac{1}{e}-\frac{e^\epsilon}{l}\Bigr)\frac{2n}{l}.
\]

We choose $l$ so that $\frac{1}{2}-\frac{1}{e}-\frac{e^{\epsilon}}{l}=\frac{1}{10}$, which yields
\[
(1-1/e)\opt-\E_{\vy\sim \mathcal{M}^{k=1}(G)}[I_G(\vy)]
\ge
\frac{1}{10e^{\epsilon}}\Bigl(\frac{4}{5}-\frac{2}{e}\Bigr)n
-\frac{4}{5}
+\frac{2}{e},
\]
which completes the proof.

\qedwhite

}

One may wonder if the lower bounds hold for approximate node- and edge-DP.  Formally, an algorithm $\mathcal{M}$ is \emph{$(\epsilon, \delta)$-node- (or edge-) DP} if for all outputs $\vy$ and node- or edge-adjacent graphs $G, G'$, $\Pr[\mathcal{M}(G) = \vy]\le e^\epsilon \Pr[\mathcal{M}(G') = \vy]+\delta$.  We often consider $\delta$ to be negligibly small, so the algorithms are approximately $\epsilon$-DP. It is not hard to see that our proof for \cref{prop:node:dp:neg,prop:edge:dp:neg} still holds for the approximated DP.  For example, 
\begin{proposition}\label{prop:epsilon-delta-hardness}
For all $n\ge 1$, $\epsilon>0$ and $\delta<1/6$, there exists $\alpha_\epsilon\ge \frac{1}{10000e^\epsilon}$ so that any $(\epsilon,\delta)$-edge-DP seeding algorithm $\mathcal{M}^{k}(G)$, there exists graph $G$ on $n$ nodes so that for $k=1$:
$\textstyle (1-1/e)\opt-\E_{\vy\sim \mathcal{M}^{k=1}(G)}[I_G(\vy)] \ge \alpha_\epsilon n.$
\end{proposition}
The proof is mostly identical to \Cref{prop:edge:dp:neg}, except that when $\mathcal{M}$ is $(\epsilon, \delta)$-DP, we only have $\Pr[\mathcal{M}^{k=1}(G)\subset S_i\cup S_j]\le {2e^\epsilon}/{l}+\delta$
.  Then the rest of the proof follows.

% \section{Proofs in Section~\ref{sec:exp} (Central DP)}\label{app:proofs:sec:exp}
\subsection{Proof of \texorpdfstring{\Cref{thm:exp:inf}}{thm:exp:inf}}\label{proof:thm:exp:inf} 

Because $\vx$ and $\vx'$ are adjacent if those matrices are different at one position, the sensitivity of the scoring function $I_\vx(v|S)$ for any $S$ is
{
\begin{align*}
    \max_{v\in [n]}\max_{\vx\underset{is}{\sim}\vx'}|I_\vx(v|S)-I_{\vx'}(v|S)| &\le \frac{n}{m}\max_{v\in [n]}\max_{\vx\underset{is}{\sim}\vx'}\left|\sum^m_{t = 1}x^t_{v}-(x')^t_{v}\right|  = \frac{n}{m},
\end{align*}} and the size of the output space is $n$.  Therefore for all $S$, \Cref{alg:dp_single} is an $\epsilon$-DP exponential mechanism. 

\textbf{Privacy:}
Let the output be $(v_1, \dots, v_k)$.  Because for any fixed $S$, \Cref{alg:dp_single} is an exponential mechanism, we have
$$\ln\frac{\Pr[\mathcal{M}_{exp}^{\epsilon/k}(\vx, S) = v_i]}{\Pr[\mathcal{M}_{exp}^{\epsilon/k}(\vx', S) = v_i]}\le \frac{\epsilon}{k} \text{ for all set }S \text{ and adjacent }\vx\underset{is}{\sim} \vx'.$$
Then we can write out the probability ratio of $\mathcal{M}_{exp}^\epsilon$ outputting the same seed set $v_1,\dots, v_k$ on two adjacent datasets $\vx$ and $\vx'$,
{
\begin{align*}
    \ln \frac{\Pr[\mathcal{M}_{exp}^\epsilon(\vx) = (v_1, \dots, v_k)]}{\Pr[\mathcal{M}_{exp}^\epsilon(\vx') = (v_1, \dots, v_k)]}
    =& \sum_i \ln\frac{\Pr[\mathcal{M}_{exp}^{\epsilon/k}(\vx, S_{i-1}) = v_i\mid S_{i-1} = (v_1, \dots, v_{i-1})]}{\Pr[\mathcal{M}_{exp}^{\epsilon/k}(\vx', S_{i-1}) = v_i\mid S_{i-1} = (v_1, \dots, v_{i-1})]} \\
    \le&  \sum_i \epsilon/k = \epsilon.
\end{align*}}
\textbf{Utility guarantee:}  Given $\vx$, we show that the output of $\mathcal{M}_{exp}^\epsilon$ is close to the optimal.  First note that once $\vx$ is fixed, \Cref{alg:original} is the classical greedy algorithm for the objective function $I_\vx$ that is submodular. Second, the error comes from the soft max of \Cref{alg:dp_single}: at each $i\in[k]$, instead of choosing the seed $v^*_i$ to maximize the score, \Cref{alg:dp_single} chooses a random seed $v_i$ according to the probabilities specified in its line \ref{iter:soft-max}.  

%\fang{made major edits, but did not affect the statements}
We set the optimal set for $I_\vx$ as $S^*:=\{v^*_1, \dots, v^*_k\}$ so that $I_\vx(S^*) = \max_{S:|S| \le k}I_\vx(S)$.  Given a seed set $v_1, \dots, v_k$, for all $i = 0,\dots, k$, let $S_i:=\{v_j:1\le j\le i\}$ and define the remainder of $S_i$ as: $R_i := I_\vx(S^*)-I_\vx(S_i) = \frac{n}{m}(|C_\vx(S^*)|-|C_\vx(S_i)|).$
Because $R_{i+1} = R_i-(I_\vx(S_{i+1})-I_\vx(S_i))$, to obtain an upper bound $R_{i+1}$, it suffices to find a lower bound on $I_\vx(S_{i+1})-I_\vx(S_i)$.  Then, because $C_\vx(S_i)\subseteq C_\vx(S_{i+1})$, $I_\vx(S_{i+1})-I_\vx(S_i) = \frac{n}{m}(|C_\vx(S_{i+1})|-|C_\vx(S_i)|) = \frac{n}{m}|C_\vx(S_{i+1})\setminus C_\vx(S_i)| = I_\vx(v_{i+1}|S_i)$.  Thus, we have
$R_{i+1} = R_i-I_\vx(v_{i+1}|S_i)$.
% Additionally, we can bound the $R_{i+1}$ by $R_i$.
% \begin{align}
%     R_{i+1} =& \frac{n}{m}\left|C_\vx(S^*)\setminus \left(C_\vx(S_{i}\cup \{v_{i+1}\})\right)\right|\\
%     =& \frac{n}{m}\left|C_\vx(S^*)\setminus C_\vx(S_i)\setminus C(v_{i+1}|S_{i})\right|\\
%     \le& \frac{n}{m}\left(|C_\vx(S^*|S_i)|-|C(v_{i+1}|S_{i})|\right)\\
%     \le& R_i-I_\vx(v_{i+1}|S_i),
% \end{align} where the second equality follows from the fact that $C_\vx(S_{i}\cup \{v_{i+1}\})$ $=$ $C_\vx(S_i)\cup C(v_{i+1}|S_{i})$. 
On the other hand, we show a lower bound of $I_\vx(v_{i+1}|S_i)$.  First, we want to apply \Cref{thm:exp:utility}.  Since the sensitivity is $n/m$, the size of the output space is $n$ for $\mathcal{M}_{exp}^{\epsilon/k}$, given $t>0$ we can set $\beta = \frac{2nk}{\epsilon m}(\ln n+t)$, and by \Cref{thm:exp:utility} and the union bound we have for all $i = 1,\dots, k$: $I_\vx(v_{i+1}| S_i)\ge \max_{w\in[n]} I_\vx(w| S_i)-\beta,$ 
with probability $k\exp(-t)$.  
Additionally, we have: 
{
\begin{align*}
    \max_w I_\vx(w|S_i)
    \ge& \frac{1}{k} \sum_{j = 1}^k I_\vx(v_j^*|S_i) = 
 \frac{1}{k}\sum_j\frac{n}{m}|C_\vx(v_j^*\cup S_i)\setminus C_\vx(S_i)|\\
    \ge& \frac{1}{k}\frac{n}{m}\left|\cup_{j = 1}^kC_\vx(v_j^*\cup S_i)\setminus C_\vx(S_i)\right|  = \frac{1}{k}I_\vx(S^*|S_i) = \frac{1}{k}R_i.
\end{align*}}
Therefore, we can combine these two inequalities and have $R_{i+1} = R_i-I_\vx(v_{i+1}|S_i)\le R_i+\beta-\max_w I_\vx(w|S_i)\le (1-1/k)R_i+\beta$.
Because $R_k = I_\vx(S^*)-I_\vx(S_k)$ and $R_0 =  I_\vx(S^*)$, the error is bounded by 
{ $$R_k\le (1-1/k)^kR_0+\sum_{i = 1}^k (1-1/k)^{k-i}\beta\le \frac{1}{e}\max_{S:|S| = k}I_\vx(S)+k\beta,$$}
where $k\beta = \frac{2nk^2}{\epsilon m}(\ln n+t)$. \qedwhite

{
\begin{lemma}[label = thm:exp:utility, name = Exponential mechanism for selecting a greedy seed]
Given $k=1$, $\epsilon>0$ and $m\in \mathbb{N}$ influence samples, $\vx$, $\mathcal{M}^{k=1, \epsilon}_{exp}(\vx)$ is $\epsilon$-ISDP and outputs a single greedy node which, for all $t>0$, satisfies:
$$\Pr\left[I_\vx(\mathcal{M}^{k, \epsilon}_{exp}(\vx))\le \max_{S:|S|=1} I_\vx(S)-\frac{2n}{\epsilon m}(\ln n+t)\right]\le e^{-t},$$ where the probability is over the randomness of the algorithm run for fixed $\vx$.
\end{lemma} %\fang{Also differentially private} \ar{I added the DP specifications to the statement of the theorem.}
This is an immediate application of Theorem~3.11 of \citet{dwork2014algorithmic}.  
Here the utility function is 
\(u(\vx,v)=I_\vx(v\mid S)=\tfrac{n}{m}\,|C_\vx(S\cup\{v\})\setminus C_\vx(S)|\),
whose sensitivity is \(\Delta(u)=\tfrac{n}{m}\).  
Plugging this sensitivity into Theorem~3.11 yields the claim.

% \fang{This is not Theorem 3.11 of \citet{dwork2014algorithmic}.  This is using exponential mechansim to find a single greedy seed.}
}

\subsection{Proof of \texorpdfstring{\Cref{thm:exp:acc}}{thm:exp:acc}}\label{proof:thm:exp:acc}

Because $m\ge {9}k(\ln n)/{\alpha^2}$, we can use \Cref{prop:inf_ch}, together with a union bound on all subsets of size at most $k$, to guarantee that $|I_\vx(S)-I_G(S)|\le {\alpha n}/{3},$ for all $S$ with $|S|\le k$ with probability at least $1-2n^k\exp(- {2\alpha^2}m/{9})\ge 1-2n^k\exp(-2k\ln n) = 1-2n^{-k}$. Therefore,
{
\begin{align}
    I_G(\mathcal{M}_{exp}^{\epsilon}(\vx))\ge I_\vx(\mathcal{M}_{exp}^\epsilon(\vx))-\frac{\alpha n}{3},\label{eq:exp_acc1}\\
    \opt = \max_{S:|S| = k}I_G(S) \le \max_{S:|S| = k}I_\vx(S)+\frac{\alpha n}{3}.\label{eq:exp_acc2}
\end{align}}
On the other hand, by \Cref{thm:exp:inf}, $\mathcal{M}_{exp}^\epsilon(\cdot)$ is $\epsilon$-ISDP.  Moreover, if we set $t = \ln n$, because $m\ge \frac{12}{\alpha\epsilon} k\ln n$, we have: 
{\begin{align*} I_\vx(\mathcal{M}_{exp}^\epsilon(\vx)) & \ge (1-1/e)\max_{S:|S| = k}I_\vx(S)-\frac{2kn}{\epsilon m}\cdot2\ln n \\ & \ge (1-1/e)\max_{S:|S| = k}I_\vx(S)-\frac{\alpha n}{3},
\end{align*}} and
{
\begin{equation}
I_\vx(\mathcal{M}_{exp}^\epsilon(\vx))\ge (1-1/e)\max_{S:|S| = k}I_\vx(S)-\frac{\alpha n}{3}, \label{eq:exp_acc3x}
\end{equation}}
with probability at least $1-\exp(-\ln n) = 1-1/n$.  
%\eqref{eq:exp_acc1}, \eqref{eq:exp_acc2},  and\eqref{eq:exp_acc3x}
Combining \cref{eq:exp_acc1,eq:exp_acc2,eq:exp_acc3x}, with probability at least $1-2n^{-k}-1/n = 1-O(1/n)$ we have:
{
\begin{align*} I_G(\mathcal{M}_{exp}^\epsilon(\vx))\ge& I_\vx(\mathcal{M}_{exp}^\epsilon(\vx))-\frac{\alpha n}{3} \tag{by \cref{eq:exp_acc1}}\\
     \ge&  (1-1/e)\max_{S:|S| = k}I_\vx(S)-\frac{2\alpha n}{3} \tag{by \cref{eq:exp_acc3x}}\\
     \ge& (1-1/e)\opt-\alpha n, \tag{by \cref{eq:exp_acc2}}
\end{align*}}
which completes the proof.

\textbf{Run time:} Note that the time complexity of computing $I_\vx(v|S)$ is $m$, which gives a $O(knm)$ bound on the time complexity of the exponential mechanism seeding algorithm, iterating the single seed selection ---  \Cref{alg:dp_single} ---  $k$ times.  \qedwhite

% \section{Proofs in Section~\ref{sec:rr} (Local DP)}
\subsection{Proof of \texorpdfstring{\Cref{prop:feasible}}{prop:feasible}}\label{proof:prop:feasible}

Suppose $C = C_{\rho,l}$ is singular for some $\rho$ and there exist two distinct vectors $\vv, \vu\in \R^{l+1}$ such that $C\vv = C\vu$.  In this proof, we will use $a$ and $b$ as indices varying from $0$ to $l$. We now assume that $\vv = (v_0,\dots, v_l)$ and $\vu= (u_0,\dots, u_l)$ are two distinct distributions such that $v_b, u_b\ge 0$ for all $b$ and $\sum_b v_b = \sum_b u_b = 1$.  We will relax this assumption later.  Because $C$ is a likelihood function where $C(a,b) = \Pr[\tilde{z}_S = a|z_S = b] = \Pr[Bin(b, 1-\rho)+Bin(l-b, \rho) = a]$ the product $Cv\in \R^{l+1}$ is also a distribution.  
Now we compute the moment-generating function of a random variable $y$ with distribution $Cv$.  Given $0\le a\le l$, let $y_b := Bin(b, 1-\rho)+Bin(l-b, \rho)$ be the random variable of the sum of two independent binomial random variables.  Then the moment-generating function of $y_b$ is $\E[e^{ty_b}] = (\rho+(1-\rho) e^t)^b(1-\rho+\rho e^{t})^{l-b}$.  Moreover, because the random variable $y$ with distribution $Cv$ is the mixture of $y_b$ with weight $v_{b}$, $y = \sum_{b = 0}^lv_b y_b$, the moment generating function of $y$ is
{ \begin{align*}
    \textstyle \E_{y\sim C\vv}[e^{ty}]
    =& \sum_{b = 0}^{l} v_{b}\E[e^{t y_{b}}] = \sum_{b = 0}^l v_{b} (1-\rho+\rho e^t)^b(\rho+(1-\rho) e^{t})^{l-b}\\
    =& (\rho+(1-\rho) e^{t})^{l}\sum_{b = 0}^l v_{b} \left(\frac{1-\rho+\rho e^t}{\rho+(1-\rho) e^{t}}\right)^b,
\end{align*}} 
for all $t$.  Similarly, the moment generating function of a random variable with distribution $C\vu$ is 
{$$\textstyle \E_{y\sim C\vu}[e^{ty}] = (\rho+(1-\rho) e^{t})^{l}\sum_{b = 0}^l u_{b} \left(\frac{1-\rho+\rho e^t}{\rho+(1-\rho) e^{t}}\right)^b.$$}
Because $C\vu = C\vv$, the moment generating functions are equal, $\E_{y\sim C\vv}[e^{ty}] = \E_{y\sim C\vu}[e^{ty}]$ for all $t$.  Since $(1-\rho+\rho e^{t})^{l}>0$ for all $t$, we have
{$$\sum_{b = 0}^l (v_{b}-u_{b}) \left(\frac{\rho+(1-\rho) e^t}{1-\rho+\rho e^{t}}\right)^b = 0 , \forall t.$$}
Because $\rho\neq 1/2$, $\frac{\rho+(1-\rho) e^t}{1-\rho+\rho e^{t}}\neq \frac{\rho+(1-\rho) e^{t'}}{1-\rho+\rho e^{t'}}$ for all $t\neq t'$, so the degree $l$ polynomial $\sum_{b = 0}^l (v_{b}-u_{b})x^b$ has more than $l$ zeros.  By the fundamental theorem of algebra, $\vv = \vu$.

Finally, suppose that there exists $\vv\neq \vu$ with $C\vv = C\vu$ but $\vv$ is not a distribution.  We show that there exist two distributions $\vv'\neq \vu'$ with $C\vv' = C\vu'$ which completes the proof.  Let $\beta := \min_{b}\{v_b, u_b\}$.  Because each row of $C$ is a distribution and $\mathbf{1}^\top C = \mathbf{1}^\top$, we can set $\alpha:=\sum_b v_b = \mathbf{1}^\top \vv = \mathbf{1}^\top C \vv = \mathbf{1}^\top C\vu = \sum_b u_b$.  We can see that the sum of $l+1$ terms should be greater than the minimum times $l+1$, $\alpha-(l+1)\beta> 0$; otherwise, $v_b = u_b = \beta$ for all $b$ and $\vv = \vu$. Then we define $\vv', \vu'\in \R^{l+1}$ so that $v'_b = \frac{v_b-\beta}{\alpha-(l+1)\beta}$  and $u'_b = \frac{u_b-\beta}{\alpha-(l+1)\beta}$ for all $b$.  We can see that $\vv'$ and $\vu'$ are two distinct distributions and that $
    C\vv'  = \frac{1}{\alpha-(l+1)\beta}C\vv-\frac{1}{\alpha-(l+1)\beta}C\mathbf{1} 
     = \frac{1}{\alpha-(l+1)\beta}C\vu-\frac{1}{\alpha-(l+1)\beta}C\mathbf{1} = C\vu',$ completes the proof. \qedwhite

\subsection{Proof of \texorpdfstring{\Cref{thm:rr}}{thm:rr}}\label{proof:thm:rr}
 Because all bits of the influence samples are flipped with probability $\rho_\epsilon$, the perturbed $\tilde{\vx}$ is $\epsilon$ locally DP.  By the post-processing immunity of DP, the output of \Cref{alg:dp_rrim} is $\epsilon$-locally DP.

For the utility, by \Cref{lem:norm_bound,lem:concentrate_infl} and a union bound, for any $\delta'>0$ and $S$ with $|S|\le k$, the estimate $J_m$ is close to the influence function $I_G$
{\begin{equation}
    |J_m(S) -  I_G(S)|\le \delta',\label{eq:rr1}
\end{equation}}
with probability $1-2n^k\exp(-\frac{2m\delta'^2}{n^2V^2_{\rho_\epsilon, k}})$ over the randomness of the perturbation. 

Next, we show that during the $k$ iterations of line~\ref{iter:1} of \Cref{alg:dp_rrim}, $S\mapsto \argmax_v J_m(S\cup \{v\})$ satisfies the conditions in \Cref{lem:additive_err}. Given \cref{eq:rr1}, for each $S$ with $|S| < k$ if $u\in \argmax_v J_m(S\cup v)$ and $u^* \in \argmax_v I_G(S\cup v)$ we have:
{\begin{align*}
    I_G(S\cup u) \ge & J_m(S\cup u)-\delta'\ge J_m(S\cup u^*)-\delta'\\
    \ge& I_G(S\cup u^*)-2\delta' = \max_v I_G(S\cup v)-2\delta'.
\end{align*}}
The second inequality holds because $u \in \argmax_v J_m(S\cup v)$.  Because the $k$ iterations of line~\ref{iter:1} of \Cref{alg:dp_rrim} satisfy the condition in \Cref{lem:additive_err} with $\delta_i = 2\delta'$ for all $i = 1, \dots, k$, the output of \Cref{alg:dp_rrim} satisfies: 
{\begin{align*}
    I_G(\mathcal{M}_{loc}^{k,\epsilon}(\vx)) & > (1-1/e)\opt-\sum_{i = 1}^k(1-1/k)^{k-i}\delta_i\\
    &\ge (1-1/e)\opt-2k\delta',
\end{align*}}
with probability $1-2n^k\exp(-\frac{2m\delta'^2}{n^2V^2_{\rho_\epsilon, k}})$.  We complete the proof by setting $\delta' = \frac{\delta}{2k}$. \qedwhite

\begin{lemma}\label{lem:additive_err}
Suppose $A:2^{V}\to V$ is an iterative single seed selection algorithm with a sequence of errors $\delta_0, \delta_1,\dots, \delta_{k-1}>0$ so that for all $S\subseteq [n]$ with $|S|=i < k$, $I_G(S\cup A(S))> \max_v I_G(S\cup v)-\delta_i$.  If $S_{i} = S_{i-1}\cup A(S_{i-1})$ for all $0<i\le k$ and $S_0 = \emptyset$, then $I_G(S_{k})>(1-1/e)\opt-\sum_{i = 1}^k(1-1/k)^{k-i}\delta_i.$
\end{lemma}

%\fang{We can improve above error bound if the resulting function  $J_m$ is submodular. One possible method is \cite{https://doi.org/10.48550/arxiv.1007.3478} If $J_m(S)$ is submodular in $S$ and we know already that it is within $\delta$ of $I_G(S)$ the error bound will have better dependence on $k$ }

\noindent{\bf Proof of \texorpdfstring{\Cref{lem:additive_err}}{lem:additive_err}:} We set the optimal $k$ seeds for $I_G$ as $v_1^*$, $\dots, v_k^*$ so that $I_G(\{v_1^*, \dots, v_k^*\}) = \max_{S: |S|\le k} I_G(S) = \opt$ and let $v_1, \dots, v_k$ be the outputs of algorithm $A$ so that $S_0 = \emptyset$ and $S_i = \{v_1,\dots, v_{i}\} = S_{i-1}\cup A(S_{i-1})$ for all $1\le i\le k$.  We define the remainder of $S_i$ as 
$R_i = I_G(S^*)-I_G(S_i)$ which is the gap between the influence of the optimal seed set $S^*$ and $S_i$. Note that $R_0 = \opt$.  We want prove that 
\begin{equation}\label{eq:additive1}
    R_{i+1}\le (1-1/k)R_i+\delta_i\text{, for all }i = 0,\dots, k-1.
\end{equation}
That will complete the proof, because $R_k \le (1-1/k)^k R_0+\sum_i (1-1/k)^{k-i}\delta_i \le \frac{1}{e} \opt+\sum_i (1-1/k)^{k-i}\delta_i$ and $R_k = \opt-I_G(S_k)$, which together imply $\opt-I_G(S_k) \leq \frac{1}{e} \opt+\sum_i (1-1/k)^{k-i}\delta_i$.

%\fang{made major edits, but did not affect the statements}

Now we prove \eqref{eq:additive1}.  Similarly to the proof of \Cref{thm:exp:inf}, first by definition, we have $R_{i+1} = R_i-(I_G(S_{i+1})-I_G(S_i))$.  Second, because $S_i\subset S_{i+1}$, $I_G(S_{i+1})-I_G(S_i) = I_G(v_{i+1}|S_i)$, we have: 
$R_{i+1} = R_i-I_G(v_{i+1}|S_i).$ To lower bound $I_G(v_{i+1}|S_i)$, by the condition of the lemma, we have:
\[
I_G(v_{i+1}\mid S_i)
= I_G(S_i\cup v_{i+1}) - I_G(S_i)
\ge \max_v I_G(S_i\cup v) - \delta_i - I_G(S_i)
= \max_v I_G(v\mid S_i) - \delta_i.
\]
 Finally, because $I_G(\cdot|S_i)$ is submodular, $I_G(S^*|S_i) \le \sum_{j = 1}^k I_G(v_j^*|S_i)\le k\max_v I_G(v|S_i)$.  By definition, $R_i \le I_G(S^*|S_i)$, and we have
$\max_vI_G(v|S_i)\ge \frac{1}{k}R_i.$  We can combine the above three inequalities and prove \eqref{eq:additive1}, 
$R_{i+1}= R_i-I_G(v_{i+1}|S_i)\le R_i-\max_v I_G(v|S_i)+\delta_i\le (1-1/k)R_i+\delta_i$. \qedwhite

% \begin{theorem}
% The output of above algorithm $\tilde{S}$ is $\epsilon$-differential private and 
% $$I(\tilde{S})\ge (1-1/e)\opt-n\left(\frac{(2-1/e)\delta}{(1-\rho_\epsilon)^k}+(1-1/e)(1-1/(1-\rho_\epsilon)^k)\right)$$
% with probability $1-n^k\exp(-2m\delta^2)$.  
% \end{theorem}

% The above error bound does not vanish as the number of samples $m$ increases.  We may use a more complicated method to estimate the probability of $\Pr[S\cap x\neq \emptyset]$ from the perturbed samples $\tilde{x}$.  Given a set $S$ and a perturbed influence sample, let $\tilde{X}_S:=|S\cap \tilde{x}|$ and ${X}_S:=|S\cap {x}|$.  There is a closed form conversion between the pmf of these two random variable
% $$\begin{pmatrix}\Pr[\tilde{X}_S = 0]\\
% \Pr[\tilde{X}_S = 1]\\
% \vdots\\
% \Pr[\tilde{X}_S = |S|]\end{pmatrix} = C_\rho\begin{pmatrix}\Pr[{X}_S = 0]\\
% \Pr[{X}_S = 1]\\
% \vdots\\
% \Pr[{X}_S = |S|]\end{pmatrix}$$
% for some matrix $C_\rho$.  We can estimate the empirical distribution of $\tilde{X}_S$.  Unbiased estimator for $\Pr[X_S = 0]$.  For all $i,j$ the entry of $C_\rho$ is 
% $$C_\rho(i,j) = \Pr[\tilde{X}_S = i\mid X_S = j] = \sum_{l}\binom{j}{l}\binom{|S|-j}{i-j+l}\rho^{i-j+2l}(1-\rho)^{|S|-i+j-2l}$$
By \cref{prop:feasible}, we know that $C_{\rho, l}^{-1}$ exists for all $l$ and $\rho\in [0,1/2)$ and $J_m(S)$ is an unbiased estimate of $I_G(S)$.  The next lemma helps us to bound the error of the estimate.
\begin{lemma}\label{lem:norm_bound}
Given $1\le l\le k$, for all $C_{\rho,l}$ with $0\le \rho<1/2$ defined in \eqref{eq:c_rho},  as $\rho\to 1/2$, 
$|1-C_{\rho,l}^{-1}(a,b)| =O((1/2-\rho)^{ -l^2})$ for all $0\le a,b\le l$. Thus\footnote{Here we first fix $k$ and take $\rho\to 1/2$.  Thus, $l\le k$ are constant relative to $1/2-\rho$.}, there exists $V_{\rho, k} = O((1/2-\rho)^{-k^2})$ where $|1-C_{\rho,l}^{-1}(a,b)| \le V_{\rho,k}$ for all $l\le k$ and $a,b\le l$.%\fang{k to 2k}
\end{lemma}

\noindent{\bf Proof of \Cref{lem:norm_bound}}: To simplify the notation, we write $C = C_{\rho,l}\in \R^{(l+1)\times(l+1)}$.  We will use $a,b$ to denote the indices from $0$ to $l$.  To bound $|1-C^{-1}(a,b)|$, it is sufficient to bound $|C^{-1}(a,b)|$.

By Cramer's rule, we can compute the entries of $C^{-1}$ from the ratios of its determinant and its cofactors' determinants. Specifically, for all $a,b = 0,1,\dots, l$, the $(a,b)$ entry of $C^{-1}$ can be obtained as follows: $|C^{-1}(a,b)| = |\det(D^{b,a})/\det(C)|$ where $D^{a,b}$ is called the $(a, b)$ cofactor of $C$, which is a submatrix of $C$ formed by removing the row $a$ and the column $b$ of $C$. Therefore, it is sufficient for us to have an upper bound on $|\det(D^{a,b})|$ for all $a,b$ and a lower bound on $|\det(C)|$.  

First, because $C$ is a column-stochastic matrix, $D^{a,b}$ is a column sub-stochastic matrix.  We define a diagonal matrix $R\in \R^{l\times l}$ where $R_{\ell,\ell} = 1/(\sum_{\imath} D^{a,b}_{\imath,\ell})\ge 1$ for all $\ell$ by \eqref{eq:c_rho}.  Because $D^{a,b}R $ is a column-stochastic matrix, $|\det(D^{a,b}R)|\leq 1$ and $ |\det(D^{a,b})| \leq |\det(D^{a,b})\det(R)| = |\det(D^{a,b}R)|.$  Therefore, $|\det(D^{a,b})|\le 1$.

Now we lower-bound $|\det(C_{\rho, l})|$.  Because $C(a,b)$ is a degree $l$ polynomial in $\rho$ for all $a,b = 0,\dots, l$, $\det(C_{\rho, l})$ is a degree $l^2$ polynomial in $\rho$.  When $\rho = 1/2$, $C_{\rho, l}$ becomes singular, and $\det(C_{1/2, l}) = 0$, and $\det(C_{\rho, l})\neq 0$ for all $0\le \rho<1/2$ by \Cref{prop:feasible}. Thus, $\det(C_{\rho, l}) = \Omega((1/2-\rho)^{l^2})$ as $\rho\to 1/2$. The latter, together with the fact that $|\det(D^{a,b})| \le 1$, implies $|C^{-1}(a,b)| = |\det(D^{b,a})/\det(C)|  = O((1/2-\rho)^{-l^2})$ which completes the proof. \qedwhite

The following is a first-order approximation when $\rho\to 0$ (not used in the paper). 
\begin{lemma}\label{lem:norm_bound2}
Given $1\le l\le k$, for all $C_{\rho,l}$ with $0\le \rho<1/2$ defined in \eqref{eq:c_rho},  as $\rho\to 0$, for all $a$ 
$$C_{\rho,l}(a,b) = O(\rho^2)+\begin{cases}
    1-l\rho \text{ if } b = a\\
    (a+1)\rho \text{ if } b = a+1\\
    (l-a)\rho \text{ if } b = a-1\\
    0\text{ otherwise}
\end{cases}\text{ and }C_{\rho,l}^{-1}(a,b) = O(\rho^2)+\begin{cases}
    1+l\rho \text{ if } b = a\\
    -(a+1)\rho \text{ if } b = a+1\\
    -(l-a)\rho \text{ if } b = a-1\\
    0\text{ otherwise}
\end{cases}$$
Therefore, $|1-C_{\rho,l}^{-1}(a,b)| \le 1+l\rho\le 1+k\rho$ for all $0\le a,b\le l$.
\end{lemma}

\begin{lemma}\label{lem:concentrate_infl}
Consider any candidate seed set $S$ of size at most $k$, $\delta>0$, and $m$ perturbed influence samples with $\rho<1/2$. Then $|J_m(S) - I_G(S)|\le \delta,$ with probability at least $1-2\exp\left(-{2m\delta^2}/{(n^2V_{\rho, k}^2})\right)$ over the randomness of the perturbed samples, where $V_{\rho, k} = O((1/2-\rho)^{-k^2})$ is defined in \Cref{lem:norm_bound}.
\end{lemma}
\noindent{\bf Proof of \Cref{lem:concentrate_infl}:} Given $S$ with $|S| = l$, the value of $J_m(S)$ is a linear function of the empirical distribution of $\tilde{f}$ on the $m$ samples, so $J_m(S)$ can be written as a sum of $m$ iid random variables.  Specifically, $J_m(S) = \sum_{t = 1}^m X_t$ is a sum of $m$ iid random variable where for $t\in [m]$, $X_t = \frac{n}{m}(1-C^{-1}_{\rho, l}(0,a))$ if the intersection of the $t$-th perturbed influence sample and $S$ has size $|\tilde{x}^t\cap S| = a$. Thus, the expectation of $J_m$ is 
{\begin{align*}
    \textstyle \E_{\tilde{\vx}}[J_m(S)] &= \frac{n}{m}\sum_{t = 1}^m\sum_{a = 0}^l (1-C^{-1}_{\rho, l}(0,a)) \Pr[|\tilde{x}^t\cap S| = a]\\ 
    =& n\sum_{a = 0}^l(1-C^{-1}_{\rho, l}(0,a)) \Pr[|\tilde{x}\cap S| = a] \\
    &= n(1-\Pr[|x\cap S| = 0])    \tag{by definition of $C_{\rho, l}$}\\
    &= n\Pr[x\cap S\neq \emptyset] = I_G(S).
\end{align*}}
Furthermore, because $|1-C^{-1}_{\rho, l}(0,a)|\le V_{\rho, k}$, $|X_i|\le \frac{n}{m}V_{\rho, k}$ by \Cref{lem:norm_bound}, we can apply Hoeffding's inequality \cite{hoeffding1994probability} to get $\Pr[|J_m(S)-I_G(S)|\ge \delta]\le 2\exp\left(-{2m\delta^2}/{(n^2V_{\rho, k}^2})\right),$ for all $\delta>0$, which completes the proof. \qedwhite
\subsection{Proof of \texorpdfstring{\Cref{thm:rr:acc}}{thm:rr:acc}}\label{proof:thm:rr:acc}

Given $\alpha>0$, we set $\delta =  \alpha n$ and $m^* = (\gamma^* {k^3V_{\rho_\epsilon,k}^2\ln n})/{\alpha^2}$ for some $\gamma^*>0$.  By \Cref{thm:rr}, $I_G(\mathcal{M}_{loc}^{k,\epsilon}(\vx))>(1-1/e)\opt-\alpha n,$ with probability 
{$$1-2n^k\exp\left(-\frac{m^*\alpha^2n^2}{2k^2n^2V_{\rho_\epsilon, k}^2}\right)\ge 1-2n^k\exp\left(-\frac{\gamma^* k \ln n }{2}\right) = 1-o(1),$$} when the constant $\gamma^*$ is large enough.

\textbf{Run time:} Note that the time complexity of computing $\tilde{\vf}$ in Line \ref{iter:empirical-dist} of \Cref{alg:offset_perturb} is $O(mn)$ and the linear system in Line \ref{iter:postprocessing} can be solved using $O(k^3)$ arithmetic operations, so computing $J_m(S_{i-1}\cup \{v\})$ in Line \ref{iter:1} of \Cref{alg:dp_rrim} takes $O(k^3 + nm)$ time. The $n k$ calls to \Cref{alg:offset_perturb} by \Cref{alg:dp_rrim}, bring the overall run time to $O(nk^4 + k n^2 m)$. \qedwhite

\section{Erd\H{o}s–R\'{e}nyi Simulation Experiments and Estimation Details}
\subsection{Simulations on Synthetic Random Graph Models}\label{sec:app:synthetic}

\begin{figure}[hbt]
\centering
\includegraphics[width=0.6\textwidth]{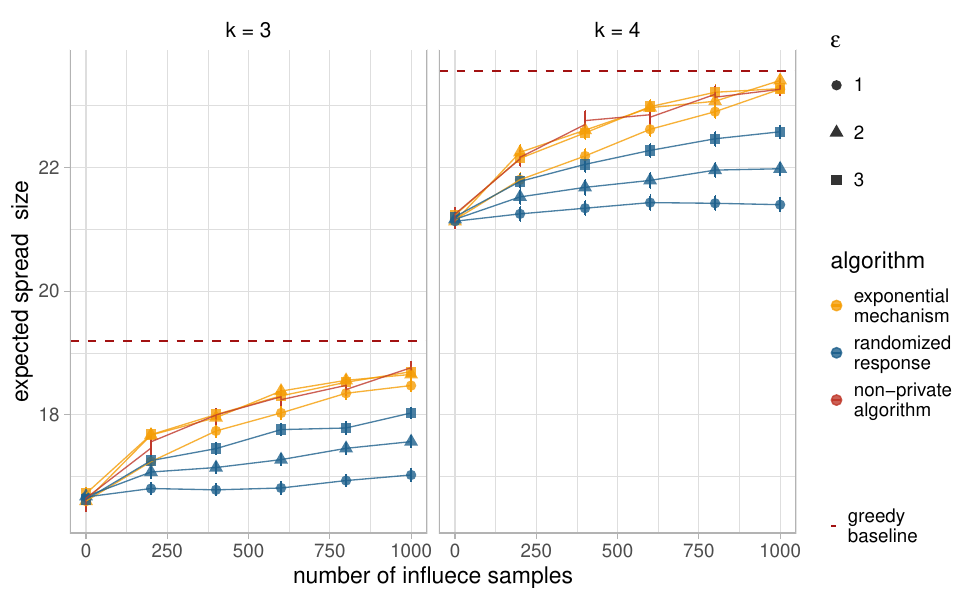}
\caption{\small 
Performance of our exponential-mechanism and randomized-response algorithms compared with the non-private influence-sample seeding method and the complete-information greedy baseline on an Erd\H{o}s–R\'enyi graph with $n=200$ and $p=0.15$. 
We report the expected influence $I_G(\cdot)$ across privacy budgets ($\epsilon$) as the number of influence samples ($m$) increases. 
Each point represents the mean over $1000$ trials, obtained from $50$ independent collections of $m$ influence samples, with each collection evaluated $20$ times. 
Error bars show $95\%$ confidence intervals, and the expected spread is estimated using an additional $1000$ influence samples.}
\label{fig:erdos-renyik24_exp_RR}
\end{figure}

To test the performance of our algorithms in a sanitized environment, we generate an input using the Erd\H{o}s-R\'enyi random graph model on $200$ nodes with edge probability $0.15$ and run the randomized response algorithm and the exponential mechanism with different privacy budgets ($\epsilon = 1, 2$, or $3$) and different numbers of influence samples ($m$ ranging from zero to $1000$). The independent cascade probability is set to $0.03$. Figure \ref{fig:erdos-renyik24_exp_RR} shows that the increase rate with $m$ slows with decreasing privacy budget for both the exponential mechanism and the randomized response algorithm.

As shown in \Cref{fig:erdos-renyik24_exp_RR}, both the exponential mechanism and the randomized response algorithm exhibit improved performance as the privacy budget $\epsilon$ and the number of influence samples $m$ increase. When $\epsilon$ is small, the injected noise limits the effective use of network information, resulting in lower expected spread. As $\epsilon \to \infty$, both mechanisms approach the non-private influence-sample greedy baseline, illustrating the trade-off between privacy protection and seeding performance.

\subsection{Estimation of STERGM parameters for MSM sexual activity networks}\label{sec:app:stergm-estimation}
Using ARTnet data, we calculated degree distributions, assortative mixing by age and race, partnership durations by age and race matching, and the probability of concurrent partnerships. These quantities were computed separately for main, casual, and one-time relationships, and we fit a distinct separable temporal ERGM (STERGM) to each. Combining simulated sexual activities from the three STERGMs allowed us to construct MSM sexual activity cascades, which we then used as influence samples for differentially private targeting in \cref{sec:msm}. In an STERGM, the transition from a network \(Y^{t-1}\) to \(Y^{+}\) is modeled through independent formation and dissolution processes, each specified by an exponential random graph model. To fit the STERGMs for the ARTnet data, we use the following formation model:
\begin{align}
    &\Pr[Y^{+}=y^{+}|Y^{t-1};\bm{\theta}^{+}] \propto \label{eq:stergm-formation-model}\exp\left[\theta^{+}_{e}e + \bm{\theta}^{+}_{d_{a}} \cdot \mathbf{d}_{a} + \bm{\theta}^{+}_{m_{a}} \cdot \mathbf{m}_{a} + {\theta}^{+}_{d_{W}} \cdot {d_{W}} + {\theta}^{+}_{m_{R}} \cdot {m_{R}} + \theta^{+}_{c}c + \bm{\theta}^{+}_{d_{cn}} \cdot \mathbf{d}_{cn}\right],
\end{align} We use bold letters for vectors. Let $e$ denote the total number of partnerships; $\mathbf{d}_{a}$ the mean degree for age group $a$ (15--24, 25--34, 35--44, 45--54, 55--64), with the 15--24 group omitted to avoid multicollinearity when estimating $\bm{\theta}^{+}_{d_a}$; and $\mathbf{m}_{a}$ the number of partnerships among individuals in age group $a$. The ARTnet data include two race categories (Black/Hispanic and White/Other), and to prevent multicollinearity we model only one, using $d_{W}$ for the mean degree among White/Other individuals and $m_{R}$ for same-race partnerships. The term $c$ counts individuals with concurrent partnerships. Cross-type degrees $\mathbf{d_{cn}}$ capture the relationship between main and casual partnerships (e.g., individuals with two casual partners have on average 0.36 main partners). To limit cross-type influence, we cap the effect of casual partnerships on main-partnership degree at 3, and the effect of main partnerships on casual-partnership degree at 2. We focus on longer-term relationships and exclude one-off effects on main and casual partnerships (which appear as “–’’ in \Cref{tab:simple_linear_model_with_robust_std_clustered_error-eastern-united-state}). Based on the formation model in \cref{eq:stergm-formation-model}, we fit STERGMs to the population-level network statistics from the egocentric ARTnet sample. Dissolution parameters are derived from partnership durations, and, following \cite{carnegie2015approximation}, we approximate formation parameters by subtracting these dissolution estimates from the ERGM estimates, reducing computational burden for large sparse networks. Parameter estimates for the main, casual, and one-off formation models are shown in \Cref{tab:simple_linear_model_with_robust_std_clustered_error-eastern-united-state}.

\begin{table}[!htbp] 
\centering 
\caption{Parameter Estimates for Main, Casual, and One-off Models}  
\label{tab:simple_linear_model_with_robust_std_clustered_error-eastern-united-state} 
\tiny
\begin{tabular}{@{\extracolsep{5pt}}lclccc} 
\toprule 
\textbf{Variable} & \textbf{Meaning} & \textbf{Main} & \textbf{Casual} & \textbf{One-off} \\ 
\hline \\[-1.8ex] 
$\theta^{+}_e$ & Total number of partnerships & -1.4327$^{\ddag}$ (0.5301) & -12.3419$^{\ddag}$ (0.7587) & -17.0396$^{\ddag}$ (2.5590) \\ 
$\theta^{+}_{d_{2}}$ & Mean degree of individuals aged 25--34 & -1.3258$^{\ddag}$ (0.2866) & 0.4676$^{*}$ (0.2536) & 3.6471$^{\ddag}$ (0.2119) \\ 
$\theta^{+}_{d_{3}}$ & Mean degree of individuals aged 35--44 & -1.0929$^{\ddag}$ (0.2819) & 1.5004$^{\ddag}$ (0.2224) & 4.2618$^{\dag}$ (1.3010) \\ 
$\theta^{+}_{d_{4}}$ & Mean degree of individuals aged 45--54 & -1.3546$^{\ddag}$ (0.2844) & 2.2343$^{\ddag}$ (0.2215) & 4.5385$^{\ddag}$ (1.2963) \\ 
$\theta^{+}_{d_{5}}$ & Mean degree of individuals aged 55--64 & -1.9009$^{\ddag}$ (0.2918) & 2.6267$^{\ddag}$ (0.2294) & 4.5570$^{\ddag}$ (1.2855) \\ 
$\theta^{+}_{d_{W}}$ & Mean degree of White/Other individuals & -1.3384$^{\ddag}$ (0.1636) & 0.0485 (0.1329) & 0.3123 (0.3561) \\ 
$\theta^{+}_{m_{1}}$ & Partnerships among individuals aged 15--24 & 0.8243$^{\ddag}$ (0.3111) & 3.7552$^{\ddag}$ (0.3465) & 9.3462$^{\ddag}$ (2.5612) \\ 
$\theta^{+}_{m_{2}}$ & Partnerships among individuals aged 25--34 & 2.2011$^{\ddag}$ (0.3686) & 1.4406$^{\ddag}$ (0.2658) & 1.8722$^{\dag}$ (0.5792) \\ 
$\theta^{+}_{m_{3}}$ & Partnerships among individuals aged 35--44 & 1.5325$^{\ddag}$ (0.3544) & 4.1681$^{\ddag}$ (0.4402) & 0.3419 (0.5552) \\ 
$\theta^{+}_{m_{4}}$ & Partnerships among individuals aged 45--54 & 1.5553$^{\ddag}$ (0.3903) & -0.5821$^{\ddag}$ (0.2834) & -0.4781 (0.5723) \\ 
$\theta^{+}_{m_{5}}$ & Partnerships among individuals aged 55--64 & 1.7164$^{\ddag}$ (0.4268) & -1.7033$^{*}$ (0.3133) & -0.9140 (0.6603) \\ 
$\theta^{+}_{m_{R}}$ & Partnerships among individuals of the same race & 0.6973$^{\ddag}$ (0.1482) & -0.0871 (0.1517) & -0.3068 (0.3939) \\ 
$\theta^{+}_c$ & Number of nodes with concurrent relationships & -3.1747$^{\ddag}$ (0.2643) & 0.4688$^{\ddag}$ (0.1413) & $-$ \\ 
$\theta^{+}_{d_{cn1}}$ & Average cross-network degree (alternative degree=1) & -0.9636$^{\ddag}$ (0.1615) & 0.9935$^{\ddag}$ (0.3172) & $-$ \\ 
$\theta^{+}_{d_{cn2}}$ & Average cross-network degree (alternative degree=2) & -1.3927$^{\ddag}$ (0.2006) & 0.4984 (0.3189) & $-$ \\ 
$\theta^{+}_{d_{cn3}}$ & Average cross-network degree (alternative degree=3) & -1.7150$^{\ddag}$ (0.2493) & $-$ & $-$ \\  \\[-1.8ex]
\bottomrule
\textit{Note:}  & \multicolumn{4}{r}{$^{*}$p$<$0.1; $^{\dag}$p$<$0.05; $^{\ddag}$p$<$0.01} \\ 
\end{tabular} 
\end{table}

We estimated the generative parameters of sexual behavior for the three partnership types (main, casual, and one-off) using separate STERGMs informed by the previously calculated network measures. These models were then used to generate influence samples that trace direct and indirect sexual relationships over time—reflecting realistic settings where such samples are observable, while the full underlying network is only hypothetical and impractical (and undesirable) to collect. In our simulation of 1,000 individuals over 120 weekly periods, grouped into 10 quarters, we selected 150 initial nodes per quarter and identified all individuals directly or indirectly connected through sexual activity, producing 1,500 influence samples. A second set of 1,500 samples, generated by choosing different initial nodes, served as a test set for evaluating the seed sets produced by our DP seeding algorithms.

{\subsection{Privacy Semantics and Extensions of ISDP}\label{sec:app:ISDP_Ext}

To clarify the privacy notion in our framework, ISDP protects \emph{activity-level} participation—whether an individual takes part in a specific diffusion event. Each influence sample represents one observed cascade and records who becomes involved; aggregating such samples yields a dataset of individuals’ participation in observable diffusion activities. ISDP therefore safeguards these participation records rather than structural connections in an underlying social network. In contrast, traditional node- or edge-level differential privacy assumes a fully known network whose nodes or edges can be perturbed, which is unrealistic in practice and would require adding far more noise, severely degrading performance. Although one could in principle adapt our exponential mechanism to satisfy node- or edge-level DP by scaling the noise from $\epsilon$ to $\epsilon/m$, this would substantially reduce utility.

%\fang{and individuals can rarely change the network of contagions.}

% Even if the underlying network were known, constructing a DP algorithm on it could drastically alter the diffusion process.\fang{I am not sure node DP need to change the graph. For instance, our algorithm with larger noise is node DP.} For instance, in a star graph, removing the central node halts all spread—changing every entry in the influence-sample matrix. \fang{Below we discuss how to modify our exponential mechanism to achieve node and edge differential privacy using larger noise $\epsilon$ to $\epsilon/m$.  (remove the rest)}

% \vspace{-1em} 
% \begin{proposition}[label = {prop:reduction}, name = {Reduction to Node- and Edge-Level DP}]
% Consider an exponential mechanism $\mathcal{M}$ that satisfies $\epsilon$-\emph{influence-sample differential privacy} ($\epsilon$-ISDP). 
% If the same mechanism is required to satisfy \emph{node-level} or \emph{edge-level} differential privacy, 
% then its effective privacy budget must be reduced to $\epsilon/m$, 
% where $m$ denotes the number of influence samples.\fang{I rewrite the proposition below}
% \end{proposition}
% \vspace{-1em} 

\begin{proposition}[label={prop:reduction}, name={Reduction to Node and Edge DP}]
Given $k \le n \in \mathbb{N}$, $\epsilon>0$, and $m \in \mathbb{N}$ influence samples $\vx$, 
$\mathcal{M}^{k,\epsilon}_{\mathrm{exp}}(\vx)$ in~\cref{alg:dp_single} is 
$\frac{\epsilon}{m}$-node DP and $\frac{\epsilon}{m}$-edge DP.
\end{proposition}

\proof{Proof of \texorpdfstring{\Cref{prop:reduction}}{prop:reduction}}
Let $\Phi:\mathbb{G}\!\to\!\mathcal{X}$ map a network $\mathcal{G}\in\mathbb{G}$ to its $m$ influence samples $\vx=\Phi(\mathcal{G})$. 
Under ISDP, two datasets $\vx,\vx'$ differ in one individual’s participation in one sample, whereas under node- or edge-level DP, $\mathcal{G},\mathcal{G}'$ differ by a single node (and its incident edges) or a single edge.
For the exponential mechanism with utility $u(\vx,v)=\tfrac{n}{m}|C_{\vx}(S\!\cup\!\{v\})\setminus C_{\vx}(S)|$ for $v\in[n]\setminus S$, the global sensitivity $\Delta(u)=\max_{Z\sim Z'}\max_v|u(Z,v)-u(Z',v)|$ satisfies $\Delta_{\text{ISDP}}(u)=n/m$ because changing one entry in $\vx$ affects at most one sample. 
In contrast, modifying one node or edge can alter the outcomes of all $m$ samples (e.g., in a star graph), yielding $\Delta_{\text{node/edge}}(u)=n$. 
Since the exponential mechanism guarantees $\epsilon$-DP with exponent $\exp(\epsilon u/2\Delta(u))$, maintaining the same privacy level under node- or edge-level adjacency requires $\epsilon_{\text{node/edge}}=\epsilon_{\text{ISDP}}\Delta_{\text{ISDP}}(u)/\Delta_{\text{node/edge}}(u)=\epsilon_{\text{ISDP}}/m$. 
Hence, an $\epsilon$-ISDP mechanism satisfies node- or edge-level DP only with effective budget $\epsilon/m$. \qedwhite
\endproof
%\fang{
Note that to achieve $\epsilon$-node or edge-level DP, we need to run $\mathcal{M}^{k, \epsilon/m}_{exp}$ and the upper bound of utility loss $(1-e^{-1})I_\vx(S^*)-I_\vx(\mathcal{M}^{k, \epsilon/m}_{exp}(\vx)) \le 2k^2n(\ln n+t)$ is independent of the number of influence sample $m$ by \cref{thm:exp:inf}.  This align with \cref{prop:node:dp:neg,prop:edge:dp:neg} because node- or edge- DP's protection should be invariant under the number of influence samples.  On the other hand, increasing the number of influence sample only reduce the estimation error between $I_\vx(S^*)$ and $I_\mathcal{G}(S^*)$ but does not shrink the DP‑induced optimization gap.  Finally, while mechanisms other than \cref{alg:dp_single} may be used for node‑level or edge‑level privacy, \cref{prop:node:dp:neg,prop:edge:dp:neg} imply that the same utility gap persists.%}

Another natural extension of ISDP is \emph{individual-level ISDP}, where individuals may participate in multiple cascades—for example, frequent contacts or high-risk participants. In this setting, changing one individual’s participation affects several samples, corresponding to a full column in the influence-sample matrix. To guarantee privacy across cascades, the privacy budget $\epsilon$ is conservatively scaled to $\epsilon/m$, where $m$ denotes the number of influence samples. This scaling is a worst-case bound, since in practice a person rarely appears in all $m$ cascades; the noise magnitude can therefore be adjusted based on empirical participation frequencies.

Overall, ISDP captures privacy at the level of observable diffusion activities rather than latent network structures, providing a realistic and flexible framework for privacy protection while preserving both analytical tractability and algorithmic utility.

}